# Smart copolymer microgels with high volume phase transition temperature: Composition, swelling, and morphology

Aditi Gujare[1,2$], Stefanie Uredat[1,2$], Jonas Runge[1,2], Felix Morgenstern[2], Domenico Truzzolillo[1], Thomas Hellweg[2]*, Julian Oberdisse[1]*

[1] Laboratoire Charles Coulomb (L2C), University of Montpellier, CNRS, 34095 Montpellier, France.
[2] Department of Physical and Biophysical Chemistry, Bielefeld University, Universitätsstr. 25, 33615 Bielefeld, Germany

* Authors for correspondence : julian.oberdisse@umontpellier.fr, thomas.hellweg@uni-bielefeld.de

$ These Authors have contributed equally to this article.

## Abstract

The thermosensitivity and microstructure of microgels made by copolymerizing standard microgel-forming monomers with more hydrophilic comonomers is investigated, with the aim of increasing the volume phase transition temperature (VPTT). We precisely determine the incorporation of N-(hydroxymethyl)acrylamide (HMAM) and purpose-synthesized N-(2-hydroxyisopropyl)acrylamide (HIPAM) into microgels—neither of which forms microgels on its own by precipitation polymerization. The swelling properties and microstructure of the resulting copolymer microgels with N-isopropylacrylamide (NIPAM, LCST ca. 32°C) and N-isopropylmethacrylamide (NIPMAM, LCST ca. 44°C) are then characterized via turbidimetry, DLS, and AFM. At low comonomer contents, all microgel particles exhibit moderate growth in size. Beyond a system-specific threshold, we observe a significant jump in size, and smoother swelling behavior. For NIPAM-HIPAM, the size increase is linked to a strong rise in swelling capacity, and the formation of a thick corona. The effect of the hydrophilic comonomers on the VPTT correlates linearly with their true composition, allowing us to extrapolate the VPTT of hypothetical pure HMAM and HIPAM microgels. This leads to 99°C for HMAM, and 68°C for HIPAM for the respective VPTT. These numbers can be seen as useful indicators of the effect of these monomers on the VPTT in the copolymerized microgels. The observed changes in VPTT, swelling, size, and morphology suggest that high-VPTT microgels possess unique internal molecular composition gradients, likely due to hydrophobic interactions during synthesis. Our results have potential implications for developing temperature-sensitive microgel-based membranes that can self-adapt their permeability at higher operating temperatures in energy applications.

## Introduction

Acrylamide-based microgels are usually spherical, colloidal particles made of crosslinked polymer swollen by a solvent which is water in most cases. [1–10] The typical size of microgels is between 50 nm and 1000 nm. The so-called smart microgels react by (de)swelling to changes of one or several external stimuli. The most studied parameter is probably the temperature.[11–14] The critical temperature where a drastic change in size of the microgels occurs is called volume phase transition temperature (VPTT). This temperature is associated with a nanophase separation inside the particles. Other parameters may affect ionic interactions, such as for instance pH [15–20], which acts on the ionization state of weak acids such as acrylic acid comonomers[21], light which can change molecular conformations and thereby hydrophilicity[22,23], or the hydrostatic pressure with its impact on intermolecular interactions[24–26]. There are numerous applications for microgel particles, as summarized in review articles. [6,9,10] They include progress from delivery systems for medications[27–32], smart catalytic systems[33–35], photonic applications[36–38], smart surfaces for vertebrate cell manipulation[39], and stimuli-responsive membranes. [40,41]

Using microgel particles for membrane formation, e.g. by spraying, is particularly suitable for thin membranes on rough supports. The formation of smart membranes made of microgel particles has been recently reviewed, where "smart" refers to the control of membrane or surface properties, such as semi-permeability ("gating"), or adsorption/desorption of target molecules, by the above-mentioned external parameters. [42–44]

Depending on the application, the VPTT may be *tailored* to meet conditions required by an application. For instance, water treatment requires rather low temperatures for costs and energetical reasons [45], while drug delivery must be triggered close to the body temperature [10], even if local heating may be used. For low-temperature fuel cells, a cut-off functionality has been imagined as intrinsic security mechanism to avoid overheating, where "low-temperature" means in the 75 − 100°C range.[46–50] As opposed to the above-mentioned applications, the VPTT thus needs to be increased, approaching the boiling point of water. It is the objective of the present article to propose novel synthesis routes with special focus on high VPTTs of microgels, for later incorporation into smart membranes, and investigate the thermodynamic properties and structure of the resulting microgels.

The physico-chemical mechanism of microgel swelling is linked to the change in hydrophobicity of the polymer chains. Microgel particles in water collapse at temperatures above the VPTT and the latter depends on the type of monomer,[51] including unconventional ones,[52] as well as on deuteration employed to adjust contrast in neutron scattering.[53] It results that the VPTT of the polymer can be tuned by the balance of hydrophilic and hydrophobic properties of the monomers, and in particular by statistically copolymerizing monomers of two different hydrophilicities.[20,54–59] For instance, Krüger et al. copolymerized NIPAM with *N-tert*-butylacrylamide, a non-thermoresponsive hydrophobic monomer, and achieved lower VPTTs.[60] Similar results were obtained with isopropyl methacrylate as comonomer by Ma et al.[61]. A remarkable change of VPTT microgels and thermogels has been also reported by employing comonomers such as N-tert-butylacrylamide (NTBAM)[62–64], hydroxyethyl acrylamide (HEAM)[20,64], oligo(ethylene glycol) (OEG) chains[65] and copolymerizing PNIPAM with poly-oligo(ethylene glycol) methacrylates (POEGMA)[66]. The first one (NTBAM), being hydrophobic in nature, decreases the microgel VPTT. Conversely, adding hydrophilic monomers, like hydroxyethyl acrylamide (HEAM) or oligo(ethylene glycol) (OEG) chains, increases the critical temperature. The increase in VPTT is a result of the enhanced hydrophilicity within the polymer network, requiring a higher temperature to induce the nanophase separation. The extent of this tuning can also be controlled by varying the comonomer feed, the length of the copolymer chains, or their structural distribution within the microgel. By manipulating these factors, the transition behavior of PNIPAM-NTBAM or PNIPAM-HEAM/PNIPAM-OEG gels can be modulated for various applications. Analogously POEGMA-PNIPAM microgels show similar versatility with a VPTT that can be tuned by varying the POEGMA chemistry and molar fraction.

In addition to that, the combination of two monomers of quite different LCST, N-n-propylacrylamide (NNPAM, 20°C) and NIPMAM (44°C) allowed Wedel et al. to maintain the synthesis conditions and the size of statistical microgel particles within a given range, while varying continuously the VPTT between the two limiting values. [56] Similar results were found by Hannappel et al.[52] Furthermore, by changing the morphology, as with core-shell-particles, which can be seen as a limiting case opposite to statistical copolymerization, [58,67] new effects have been found.[68,69] Finally, as a special case of copolymerization, crosslinking molecules may also modify the VPTT.[70,71]

In this paper, we study the effect of copolymerization of a standard microgel-forming monomers N-isopropylacrylamide (NIPAM, LCST ca 32 °C[12]) and *N*-isopropylmethacrylamide (NIPMAM, LCST ca. 44 °C[67]), see also ref. [52], and a second, more hydrophilic monomer. These two comonomers, *N*-(2-hydroxyisopropyl)acrylamide (HIPAM) and *N*-hydroxymethylacrylamide  (HMAM) do not form microgels by themselves in precipitation polymerization, because of their hydrophilicity. The molecular structure of the monomers employed in the present work are shown in Scheme 1. To our knowledge, the effect of copolymerizing NIPAM with HMAM, a hydrophilic monomer of relevance for reaching higher phase separation temperature as reported in this article, has not yet been investigated in microgels. Ternary (macroscopic) hydrogels containing increasing ratios of HMAM of up to ca. 45 mol% have been synthesized by Işik and Günay. The evolution of the equilibrium swelling ratio as a function of temperature showed that the VPTT shifts from 32 °C to 35 °C for an incorporation of HMAM of 30 mol%, before widening the shape of the transition considerably, progressively losing thermosensitivity.[72] HMAM-containing copolymers – up to 20 mol% – and including a pH-trigger have been shown by Yu et al. to have LCSTs which could be varied between 30 °C and 50 °C.[73] Higher VPTTs of up to 55 °C of copolymer chains grafted from silica particles have been obtained by copolymerizing pNIPAM with up to 30 mol% of HMAM.[74] Copolymers of HMAM have been reported in the literature to form nanofibers, hydrogels, nanoparticles, micelles, and hollow spheres. [51,75–78] Finally, linear NIPAM-co-HIPAM copolymers have been synthesized by Maeda et al.[79] They obtained a maximum LCST from the cloud point of ca. 80 °C for a 80 mol%-HIPAM content. In the present paper, the true incorporation rate of the comonomers has been determined by IR and NMR to obtain a quantitative analysis of the impact of HIPAM (resp. HMAM) on NIPMAM or NIPAM systems, and correlate it with swelling properties and morphology. The volume phase transitions have been investigated by DLS and turbidimetry, and the microgel morphologies by AFM.

## Materials and methods

**Monomers and chemicals:** The four monomers used in this study are depicted in Scheme 1. The main monomers were N-(isopropyl)acrylamide (NIPAM, BLDpharm, 99.82%, or Sigma-Aldrich, >99 %), N-isopropylmethacrylamide (NIPMAM, Sigma Aldrich, 97%), and the crosslinker N,N'-methylen bisacrylamid (BIS, from Sigma Aldrich, 99 %). Details on concentrations are given in the SI. One comomonor, N-(2-hydroxyisopropyl)acrylamide (HIPAM) has been synthesized in our lab according to Maeda et al.,[79] and its molecular structure characterized by $^1$H-NMR (Figure S1 in the SI). The second comomonor N-(hydroxymethyl)acrylamide (HMAM) has been purchased from Sigma Aldrich (48 wt % in H$_2$O). The initiators were ammonium persulfate (APS, from Roth, >98 %;), potassium persulfate (KPS, from Sigma Aldrich 99%).

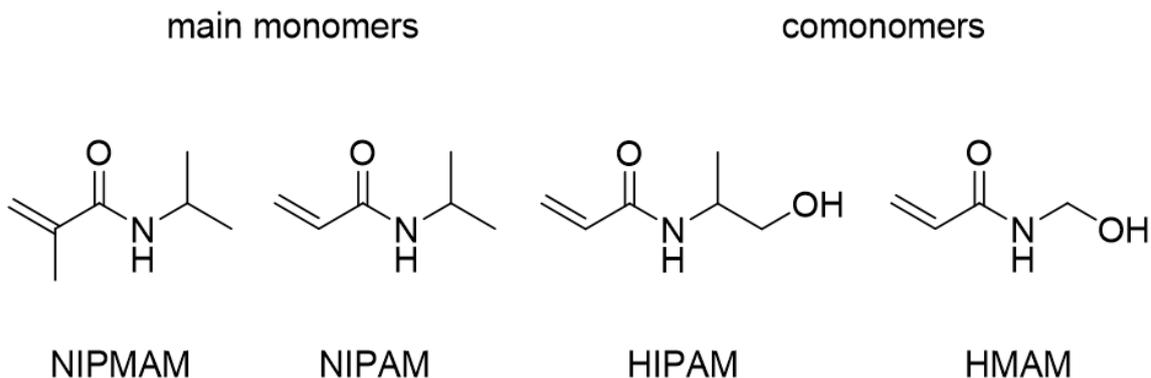

**Scheme 1:** Structures of the monomers used in the microgel synthesis.

**Microgel synthesis:** The NIPAM and NIPMAM monomers are used to copolymerize the more hydrophilic monomers HIPAM and HMAM to increase the VPTT of the microgels. The microgels were synthesized via surfactant-free precipitation polymerization.[80] The monomers and the crosslinker BIS (5 mol% of the total monomers) were dissolved in distilled water (~1 wt. % solution) under inert gas conditions followed by the addition of a persulfate initiator (KPS/APS, 1 mol% of the total monomers, at a temperature of 80-90°C, well above the expected volume phase transition temperature (see Table S1). The reactions were allowed to proceed under stirring for at least 2 hours at 400 rpm at this temperature after which the heating was stopped. The reaction mixture was allowed to cool down to room temperature and was stirred overnight. Purification was done by three successive centrifugation (at room temperature), decantation, and redispersion cycles. The compositions of the microgels are described in Table S2 and S3, and comparisons between slightly varying synthesis routes are shown in Figures S2 and S3.

The sample names encode the nominal composition in main monomer and comonomer (with fixed BIS which is always present), e.g. NIPAM-HMAM 80:20 correspond to 20 mol% of HMAM out of all monomers in the reaction mixture. Above some threshold in comonomer content, we were unable to synthesize microgel particles, see the example of a failed synthesis in Figure S4. Therefore, in order to investigate if the pure HMAM, resp. HIPAM molecules without crosslinker possess an LCST, linear molecules have been synthesized over 24h following Maeda et al. at T = 60°C in dimethylformamide (DMF) with APS as initiator[79] and purified via dialysis against water.

**Light scattering:** The swelling curves of the microgel particles have been measured by dynamic light scattering (DLS) providing the hydrodynamic radius $R_h$ as a function of temperature. Samples have been measured under dilute conditions *(ca. 0.0005 wt%)* in water. The data have been analyzed with the cumulant method.[81,82]

Measurements were performed in Bielefeld at temperatures between 10° to 60 °C, at a fixed scattering angle of $\Theta_s$=45°. Instrumentation included a He-Ne-Laser ($\lambda$=632.8 nm, 21 mW, HNL210L); Thorlabs Inc., ALV-6010 multiple-$\tau$ digital autocorrelator (ALV-Laser Vertriebsgesellschaft), thermostat (C25P; Thermo Haake GmbH) with Phoenix II Controller, SO-SIPD single-photon detector (ALV-Laser Vertriebsgesellschaft and Pt100-thermometer). All temperature ramps were done by changing temperature in $\Delta T = 2$ K steps and at each temperature 5 photon correlation runs for 300 s were carried out and subsequently averaged. The measurements up to 80°C were performed in Montpellier using an Anton Paar set-up (Litesizer 500, $\lambda$=658 nm) at $\Theta_s$ = 90°. For each temperature ($\Delta T = 2$ K), intensity traces were recorded 3 times for 300 s each, and the correlation functions averaged before being analyzed by the cumulant method. Analysis with CONTIN gave equivalent results.[83,84]

**Turbidity measurements:** The cloud point of dilute microgel suspensions has been determined by measuring the relative transmittance of visible light as a function of temperature. These measurements were performed using an Anton Paar set-up (Litesizer 500, λ=658 nm). Each transmission was measured in 1 K steps for 60 s, and normalized by the transmission of deionized water at the same temperature. The sample thickness was 10 mm (square Quartz cuvette; Hellma Analytics).

Both transmission and DLS hydrodynamic radii have been fitted using the following function of temperature, giving access to the VPTT.[85]

$$R_h(T) = R_0 - \Delta R_h \tanh[s(T-T_c)] + A(T-T_c) + B(T-T_c)^2 + C(T-T_c)^3 \qquad (1)$$

Here $T_c$ is the VPTT, $R_0$ is the radius of the microgel at the VPT, $\Delta R_h$ is the amplitude of the VPT, and the parameter $s$ quantifies its sharpness. Finally, the parameters A, B, and C phenomenologically capture the deviation from a pure sigmoidal behavior, including asymmetries of the swelling curves with respect to $T=T_c$.

**Atomic Force Microscopy** (AFM) has been used to investigate the shape and morphology of the dried microgel particles deposited on wafers. The measurements were performed on a Nanosurf FlexAFM setup in tapping mode with a Cantilever from BudgetSensors (Type Tap300Al-G, resonance frequency: 300 kHz, force constant: 40 N/m). For this purpose, the silicon wafers (~1 cm²) were plasma cleaned (Zepto, Diener electronic, 0.4 mbar, 100% output, 60 s, O₂) and PEI (40 µl, 0.25 wt%) spin coated. After that, the samples (40 µl) were spin coated onto the wafer and let dry overnight.

**Comonomer incorporation by attenuated total reflectance Fourier-transform infrared spectroscopy** (ATR-FTIR)[52] and **nuclear magnetic resonance** (NMR) have been used to analyze the comonomer content inside the microgels. Both methods have been tested for both high-VPTT comonomers, but it turned out that while for HMAM, the peaks identifying the comonomers were well isolated in IR-spectroscopy, this resulted inappropriate for HIPAM, presumably due to the liquid character of the pure comonomer causing concentration gradients in the monomer mixture. Therefore, ¹H-NMR had to be used. The HMAM-based microgels were freeze-dried and the FTIR measurements were performed using the diamond single reflection ATR accessory, on a Bruker Tensor 27 spectrometer with DTGS pyroelectric detector and a GLOBAR light source. The comonomer HIPAM and the HIPAM-based microgels were freeze-dried and characterized by ¹H-NMR in D₂O (Bruker Avance III 500, 500MHz). For IR, calibration curves have been constructed by IR measurements of various mixtures of comonomers without initiating the polymerization in water, followed by freeze-drying. The identification of the relevant IR-peaks for HMAM is shown in Figure S5. The comparative analysis of peak heights and areas gave equivalent results. The peak height at 1034 cm⁻¹, where only the HMAM molecules contribute has been normed to the height of the peak at 1645 cm⁻¹, where both monomers, NIPAM (resp. NIPMAM), and HMAM contribute. For NMR, for peaks shown in Figure S6, areas were systematically used, and instrumental error bars are expected to be around 2%, i.e. much lower than the dispersion due to sample preparation. The signal at 3.51 ppm is normalized to the signal of both monomers at 3.91 ppm, yielding directly the molar fraction of HIPAM in the copolymer.

## Results and discussion

**Comonomer incorporation:** Four systems have been produced by copolymerization of the two main monomers (NIPAM and NIPMAM) of comparatively low VPTT, 32°[12] and 44°C[67], respectively, with the two more hydrophilic, high-VPTT comonomers, HIPAM and HMAM. To compare the effects of the comonomers on the VPTTs of the microgels, the true incorporation rate has been measured by either IR (for HMAM) or NMR (for HIPAM). For HMAM, the result is presented in Figure 1, based on spectroscopy data and results shown in Figure S5. HIPAM NMR data and results are reported in Figure S6. The calibration of the IR results by the pure monomer mixture gives the upper linear function in Figure 1a, and the microgels the lower one. The ratio of the slopes yields the average incorporation. Within uncertainties, a linear law with a constant incorporation efficacy describes well the microgel composition. For NMR, the calibration can be directly based on a microgel spectrum, and the true vs. nominal comonomer content has been plotted in Figure 1b.

The comonomer content is found to be on average 55% of its nominal value for NIPAM-HMAM, 69% for NIPMAM-HMAM, 80% for NIPAM-HIPAM, and 48% for NIPMAM-HIPAM. In what follows, we report all measurements as a function of the true incorporation obtained by multiplying the nominal content by the incorporation efficacy. For simplicity, the names of the samples (e.g. NIPAM-HMAM 80:20, indicating the nominal molar composition) have not been changed, whereas any result for this sample are now plotted at the true composition, i.e., 20%*0.69 = 13.8% for the above example.

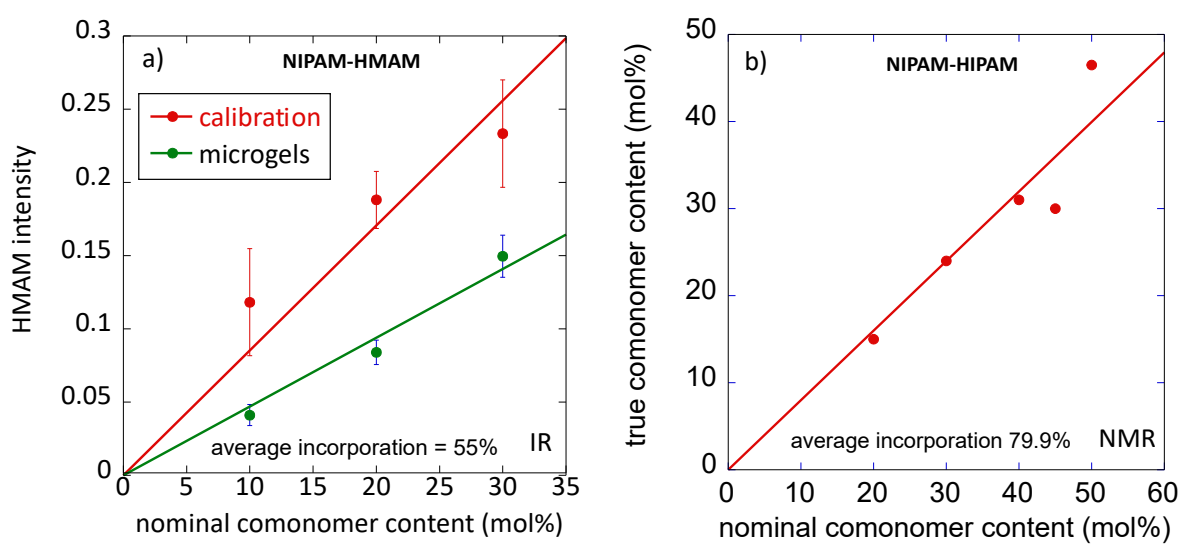

**Figure 1:** Incorporation of comonomers into the microgels. (a) IR peak intensities for the NIPAM-HMAM copolymer system as compared to the calibration, (b) NIPAM-HIPAM true comonomer content as determined by NMR. The lines are linear fits. The ratio of the slopes in a) gives the average incorporation of 55%. For NMR, it is directly given by the slope in b).

**Shift in VPTT:** The volume phase transition temperature has been determined by cloud point measurements. The transmission in water depends mainly on particle size, refractive index contrast, and on particle concentration, which had to be adapted slightly to optimize detection. The normalized transmittances of light are shown in Figure 2 for the four systems, as a function of their true comonomer content. The original transmittances are shown in Figure S7.

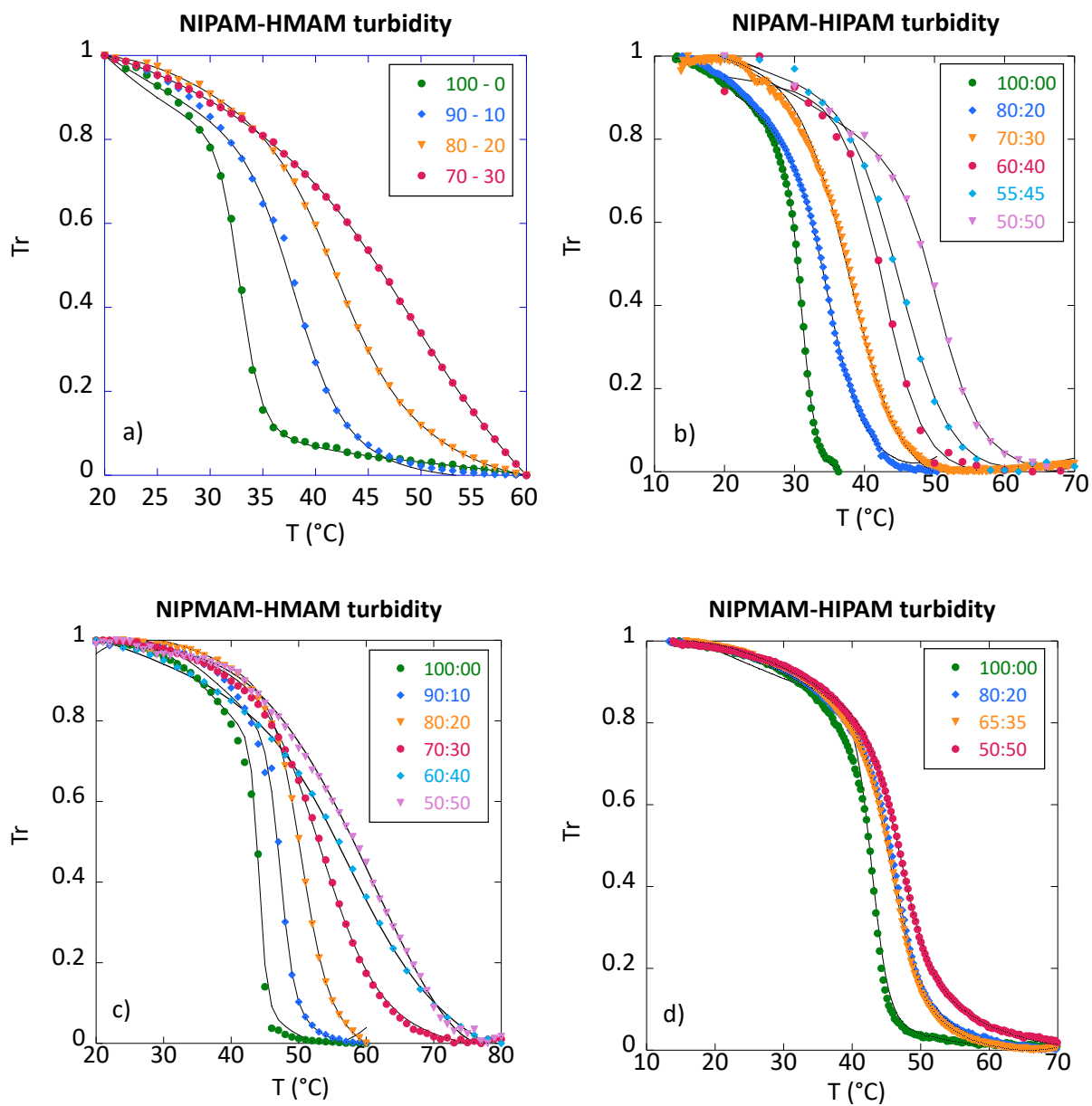

**Figure 2:** Normalized turbidity measurements reported as the transmission of dilute microgel suspensions: (a) NIPAM-HMAM, (b) NIPAM-HIPAM, (c) NIPMAM-HMAM, and (d) NIPMAM-HIPAM, as a function of temperature, for different nominal comonomer contents as given in the legends. Continuous lines are the hyperbolic tangent fit functions, see text for details.

In Figure 2, the shift in VPTT with increasing HIPAM (resp. HMAM) content emerges clearly. One system, NIPMAM-HIPAM (Figure 2d), shows a more modest shift in VPTT, which is due to the moderate incorporation efficacy of 48%: we did not succeed in incorporating more than 24 mol% HIPAM. The shape of the transmittance functions evolves from a more abrupt decay for the pure NIPAM and NIPMAM systems to a *smoother* decrease as more hydrophilic comonomer is incorporated: the volume change takes place over a larger temperature range. This is particularly true for the HMAM copolymers, as opposed to the HIPAM ones. Broader and smoother curves point to more heterogeneous structures where spatial cooperativity of the VPT is lower because of the presence of comonomers with presumably very high VPTT. These transmittances as a function of temperature have been fitted by a hyperbolic tangent function given by eq.(1) adapted for transmittance and superimposed to the data in Figure 2. The fits are generally of good quality, with a well-defined VPTT in each case. This quantity is the main outcome of these experiments, and the VPTT has been plotted as a function of the true monomer content in Figure 3. Anticipating the results of the swelling study by DLS, the VPTTs determined with the $R_h(T)$ functions have also been added on this graph. For further analysis below, both DLS and turbidity data sets have been combined in a single data set, allowing reduction of the statistical error. The two data sets are also shown separately in Figure S8. There one can see that that individual techniques deliver a rather broad dispersion of points, with some outliers strongly affecting the linear fits. On the contrary, combining the outcome of DLS and turbidity as in Figure 3 gives consistent results.

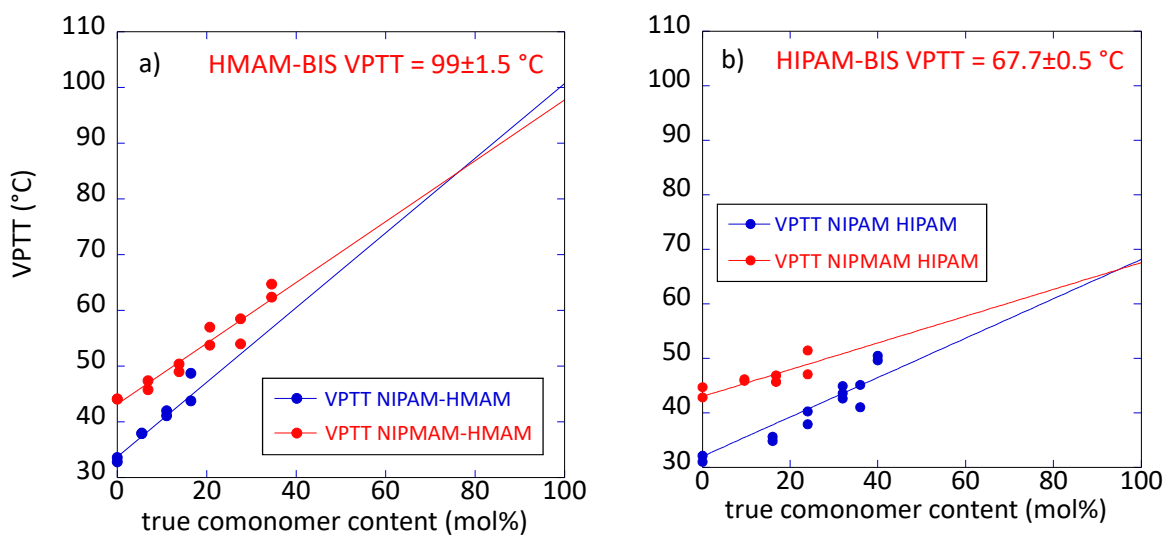

**Figure 3:** The volume phase transition temperature combining results of dynamic light scattering and cloud point measurements for **(a)** HMAM-microgel systems, and **(b)** HIPAM-microgel systems, as a function of the true hydrophilic comonomer content. All have been extrapolated to 100% hydrophilic comonomer (always including additional 5 mol% BIS), see SI for detailed compositions (Table S2 and S3).

In Figure 3, the expected VPTTs of the pure main monomer systems are found: about 32°C for NIPAM [12], and 44°C for NIPMAM [67]. Adding the hydrophilic comonomer HMAM to the latter leads to a steady increase of the VPTT, to up to more than 60°C for the highest incorporation in NIPMAM as shown in Figure 3a. For NIPAM, the transition temperatures are lower, but the increase is steeper with HMAM, due to the stronger difference in hydrophilicity between the two monomers. As indicated in Table S1, there is an upper limit of incorporation of the high VPTT comonomers into the microgels, generally around 40-50 mol%. However, both datasets seem to be well described by a linear function in Figure 3a and can be extrapolated to estimate the VPTT of a hypothetical pure HMAM (including BIS) microgel, assuming that these acrylamide molecules with alcohol function are thermosensitive. This procedure works well for the two HMAM systems, as it leads to a coherent estimation of the VPTTs of hypothetical pure microgels of hydrophilic comonomer with 5mol%BIS: (99±1.5) °C for HMAM. For the HIPAM systems, the extrapolations in Figure 3b point to a lower hypothetical VPTT of HIPAM of (67.7±0.5) °C. The VPTT of HIPAM is thus lower than the one of HMAM, and its effect on the copolymerized microgels is therefore less pronounced. Due to the limited comonomer concentration range, this extrapolation has to be taken with care. It can be seen as the prediction of strictly ideal VPTT values for pure comonomer microgels. Moreover, the limiting value of the VPTT at 100% gives a measure of the impact of a specific monomer on the VPTT in copolymerization. Moreover, it can be used as parameter of a linear law to predict the VPTT in the low-concentration range.

In order to check the thermosensitivity of the pure HMAM (resp. HIPAM) molecules, a separate synthesis of linear molecules in absence of BIS has been performed at 60°C. For comparison, linear copolymers with NIPAM have also been synthesized. Unfortunately, turbidity measurements are inconclusive: while HIPAM copolymers exhibit a composition dependent LCST, no effect was observed for pure HIPAM chains in the accessible temperature range < 80°C. For HMAM, given the similar molecular structure, thermosensitivity was expected, but no cloud point could be detected below 100°C. Miniemulsion polymerization has been identified as an alternative synthesis route, losing the benefit of the small microgel sizes, but possibly ensuring higher comonomer contents. Also, the use of an autoclave station could be an interesting alternative approach since this allows with overheated water. From the available data, the experimental fact is that the effect of copolymerizing HIPAM (resp. HMAM) with NIPAM or NIPMAM is to shift the VPTT as if the crosslinked pure HIPAM (resp. HMAM) microgels were thermosensitive, with a transition temperature of 68°C (resp. 99°C). These hypothetical transition temperatures can thus be interpreted as a parameter describing the impact of these molecules on the respective copolymer microgels.

**Swelling properties:** While the shift in VPTT can be established by cloud point measurements, the swelling properties of the four microgel systems needed to be studied by DLS as a function of temperature. The results are shown in Figure 4. As one can see, the swelling curves become quite scattered in some cases. We have analyzed the corresponding correlation functions in detail, and some of them are shown in Figures S9 and S10. To further investigate the possible presence of aggregated microgels and possible smoothing of the deswelling curves, we went beyond the cumulant analysis of the intensity correlation functions which overweights the fastest components. In addition, we have determined an average hydrodynamic radius as defined in the SI. The two methods give the same $R_h$ at low temperatures, while the average radius is slightly larger than the radius obtained via the cumulant method above the VPTT and has been plotted in Figure 4.

The hydrodynamic radius of the particles at low temperature is generally in the 200 to 400 nm range. It increases moderately with added comonomer. Above a critical hydrophilic comonomer content, an intriguing jump in size is observed in three out of four cases. The only exception is shown in Figure 4d. There is also a progressive increase in size for the remaining NIPMAM-HIPAM microgels, but no jump. Concomitantly, we observe an upturn in the deswelling curve $R_h(T)$ above the VPTT for the highest comonomer contents in NIPAM-HMAM, NIPMAM-HIPAM and NIPMAM-HMAM microgels pointing to a partial loss of colloidal stability of the microgels, presumably forming aggregates of a few microgel particles.

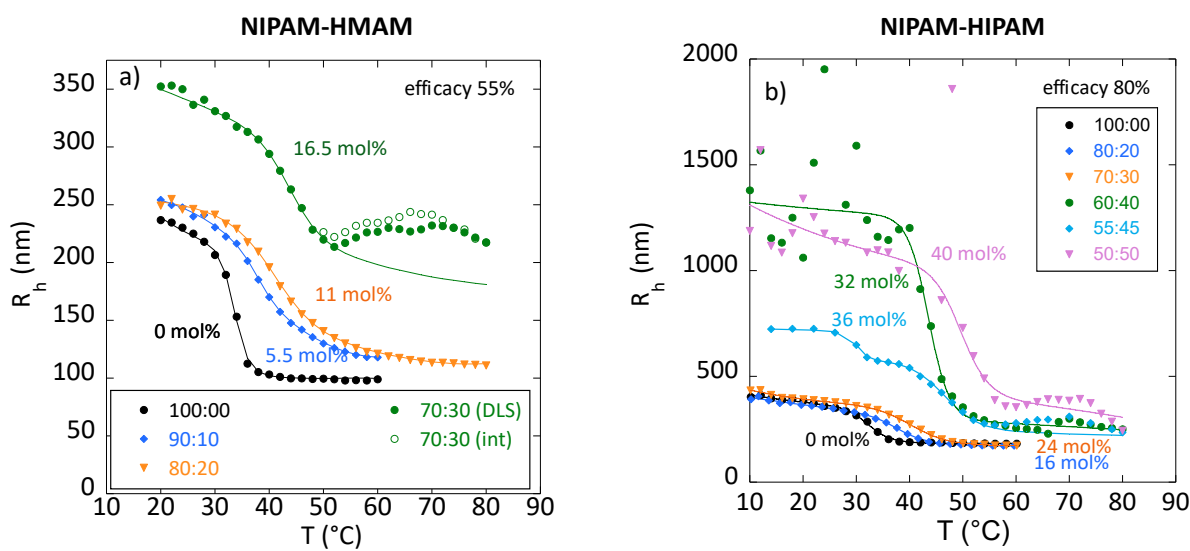

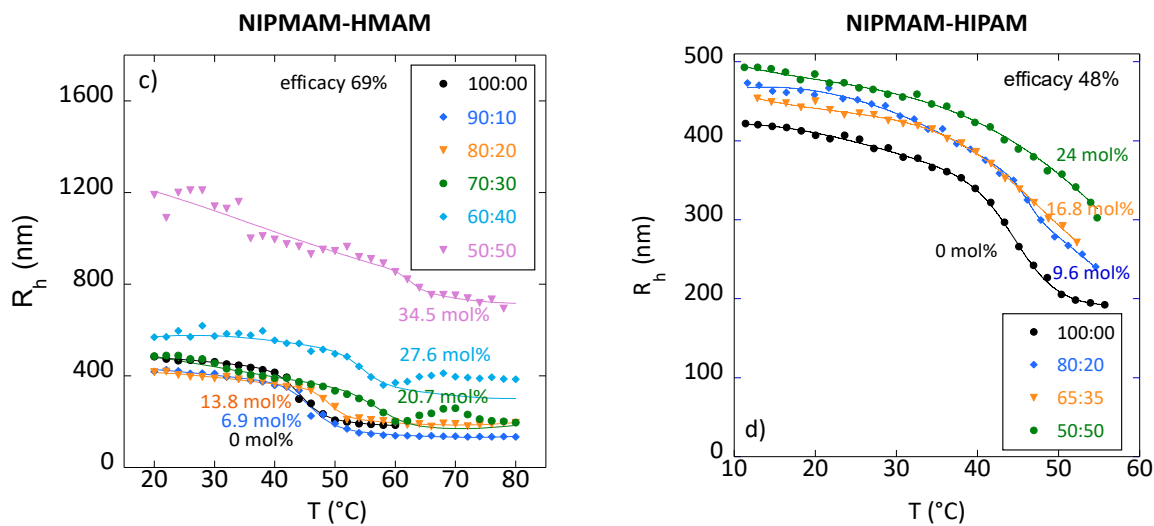

**Figure 4:** Swelling curves from the microgels systems: (a) NIPMAM-HMAM, (b) NIPMAM-HIPAM, (c) NIPMAM-HMAM, (d) NIPMAM-HIPAM, for different comonomer contents as indicated in the legend. True incorporations are given as labels. Continuous lines are fits based on eq. (1).

As soon as the microgel systems approach the abrupt jump in size, the $R_h(T)$ curve becomes more scattered. As shown in the SI, we have inspected the intensity traces recorded in DLS experiments, see Figure S10. The jump in size is systematically accompanied by an unavoidable rise of temporally localized large fluctuations of the scattered light. At the same time, a small but detectable shoulder in the correlation functions is found at long lag times. These features are the fingerprint of the formation of large objects passing through the scattering volume during the measurement. Along with the overall increase in size as determined by the cumulant method, these findings clearly indicate a loss of homogeneity in the microgel size distribution and possibly represent a precursor to a size divergence, suggesting the eventual impossibility of forming sub-micrometric microgels beyond a certain comonomer content threshold. In order to test reversibility, we have performed one heating and one cooling ramp, and have not observed any relevant hysteresis. Some aggregation is observed upon heating above the VPTT in some cases, as in Figure 4a (16.5 mol%), pointing towards loss of colloidal stability, but also this process seems to be reversible, because the low-T size goes back to its initial value after one cycle. Finally, the AFM observations discussed below do not show any large particles or aggregates, presumably due to the low statistics in terms of particle numbers. This suggests that large microgel particles are very rare, but when they cross the illuminated scattering volume, they generate the artifacts discussed above, and thus the noisy outcome of the DLS experiments. In spite of these shortcomings, the DLS swelling curves of Figure 4 convey a strong message: with increasing comonomer content, microgels grow in size, first moderately, then more abruptly, probably becoming less monodisperse. At the same time, the phase transition is smoothed, presumably as the result of the superposition of microgels of different compositions, or of a possible internal composition gradient, and most importantly, the average VPTT is shifted to larger temperatures, as already shown and compared in Figure 3 to the turbidity results.

The jump in microgel size with composition can be characterized further at low temperature, i.e. in the swollen state. In Figure 5, the average $R_h$ at 20°C of copolymerized microgels has been normed to the value of the pure microgel in absence of comonomer and plotted as a function of the true incorporation of hydrophilic comonomer for the four systems. One might have expected that the microgels behave similarly as a function of the true incorporation, but as shown in Figure 5, the microgels show a jump in size at about 15 mol% for NIPMAM-HMAM, and at around 35 mol% for NIPMAM-HIPAM and NIPMAM-HMAM. The last system, NIPMAM-HIPAM, does not reach such high incorporations, and shows no jump, but a progressive shift of its VPTT.

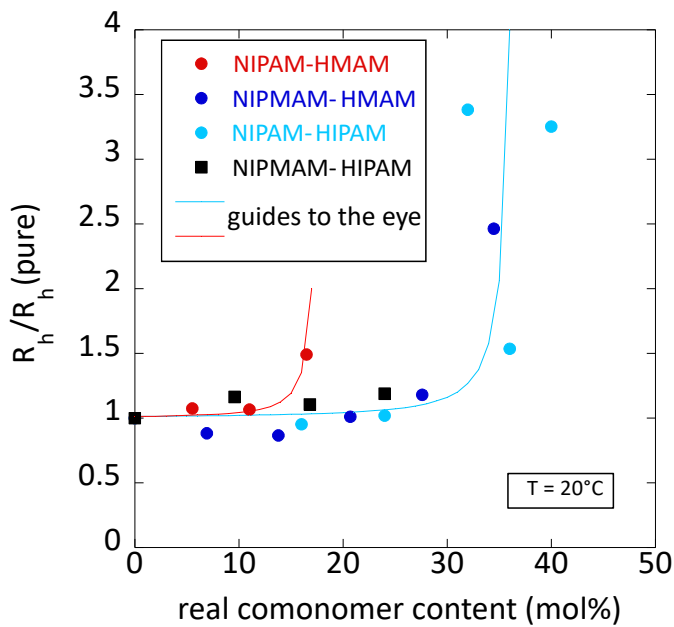

**Figure 5:** Hydrodynamic radius of swollen microgels at 20°C normalized to the ones of the pure NIPAM (resp. NIPMAM) microgels, as a function of true comonomer content as indicated in the legend. The continuous lines are a guide to the eye.

The origin of this jump remains unknown, but it can be conjectured that the presence of a too large fraction of hydrophilic monomers does not contribute efficiently to the precipitation mechanism even at high temperature. The latter has been raised to 80-90°C to favour precipitation, see Table S1. Rather, the more hydrophilic monomers may remain hydrated and gather around the few nuclei of the particles in formation, the population of which diminishes as the amount of precipitating main monomer decreases. This induces the formation of a large, less crosslinked, and thus possibly more swellable corona. In this case, the morphology may be impacted, as well as the maximum swelling ratio, and both will be investigated below. We have attempted comonomer compositions above the observed jump in size, and either no microgels were formed, resulting in a clear solution as with NIPMAM-HMAM, or large lumps of polymer were obtained for NIPAM-HMAM. For NIPAM-HIPAM, a macroscopic polymer layer forms during the synthesis.

These results suggest that the jump in size is a first manifestation of changes in molecular structure, making the precipitation polymerization of pure microgels of either HIPAM or HMAM impossible. This mechanism would be quite different from others observed in the literature. By statistically copolymerizing microgel-forming monomers such as NNPAM and NIPMAM, Wedel et al. observed a linear increase in size with higher hydrophilic monomer content.[56] These authors explained the larger sizes with a lower number of precursor particles, but the morphology of the particles remained similar. In other statistical copolymer microgels, the sizes were found to be relatively constant as compared to the large jumps observed here.[52] We believe that the strong difference in hydrophobicity in the present investigation is responsible for the strong impact on particle sizes, whereas the hydrophobicity of the comonomers studied in the cited references are more similar to the one of NIPAM.

Regardless of the change in size, the total maximum volumetric swelling ratio of microgels (as opposed to the ratio of the radii) may be affected by the hydrophilic comonomer, hinting at changes in molecular structure. Such information is contained in the low and the high T size reported in the (de-)swelling plots in Figure 4. If a plateau is reached at high temperatures, the totally collapsed size is known – if not a lower estimate for the swelling can be estimated. In Figure 6, we have represented the maximum hydrodynamic volume swelling ratio given by $(R_{max}/R_{min})^3$, as a function of true comonomer content. Here $R_{max}$ denotes the average size at 20°C, and $R_{min}$ is the smallest measured radius, both measured by the swelling curves shown in Figure 4. The NIPAM-HIPAM microgels are found to have increasing maximum swelling ratios with composition, reaching ratios as high as one hundred in a manner resembling a divergence in size as indicated by the guide to the eye superimposed to the data. This divergence is found to take place at the same comonomer content as the jump in size described in Figure 5. It corresponds to a strong increase in size at low temperatures, but a still rather small collapsed state at high temperature. As a result, the ratio between the two sizes increases strongly.

A molecular interpretation could be that a rather hydrophilic corona is formed, which extends far out into the solvent – and thus contributes strongly to the hydrodynamic radius –, but collapses completely onto the microgel core at high temperature. All other microgel systems start from a maximum volumetric swelling ratio below 20 for the pure microgels (NIPAM, resp. NIPMAM). As the comonomer ratio is increased, the maximum swelling ratio decreases for these three systems, indicating that a completely different molecular structure is favoured for these systems. Along the same hypothesis as above for the NIPAM-HIPAM system, the particles seem to be more homogeneous and more hydrophilic, with bigger collapsed sizes leading to a globally lower total volumetric swelling.

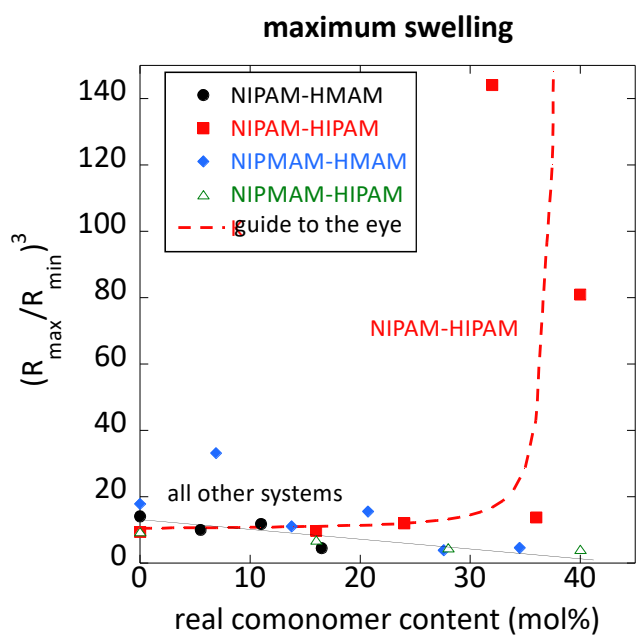

**maximum swelling**

**Figure 6:** Maximum volumetric swelling ratio of microgels between low and high temperature as described in the text (solid symbols), as a function of the true comonomer content. Partial swelling ratios are reported in case of incomplete collapse, i.e. for NIPMAM-HIPAM.

Morphologies of the copolymerized microgel particles measured with an imaging technique will be presented in the next section in order to provide additional data allowing for molecular understanding of the swelling curves.

**Morphology of microgel particles:** AFM measurements have been performed to study the microgel morphologies. The particles have been deposited on a silicon wafer and observed in the dry state. The sizes obtained by AFM cannot be directly compared to the ones observed in suspension by light scattering: they are de-swollen and flattened on the wafers. In Figure 7, representative phase images of the HMAM-rich microgels are shown. A more exhaustive set of images is provided in Figures S11 and S12. Particles are always found to have a circular cross-section, and even at high magnification, they show a homogeneous structure, with possibly a very thin corona, as shown in Figure 7d. This corona is interpreted as the usual surface of standard microgels, including necessarily some dangling ends, but no truly developed thick corona structure.

**a) NIPAM-HMAM 90:10**

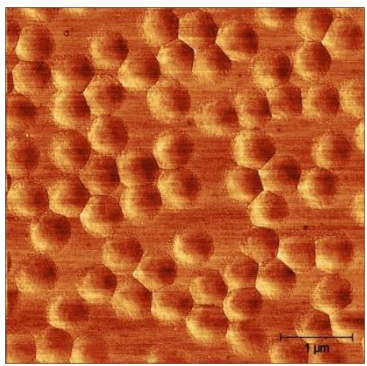

**c) NIPMAM-HMAM 80:20**

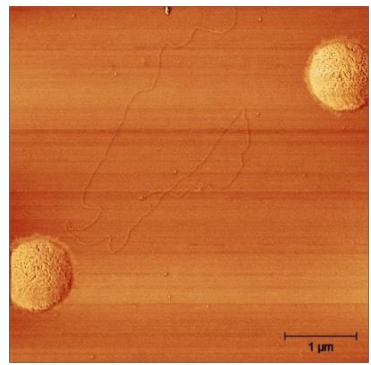

**b) NIPAM-HMAM 80:20**

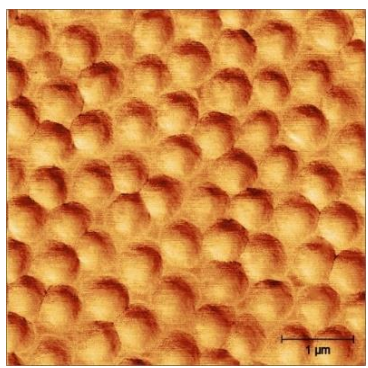

**d) NIPMAM-HMAM 50:50**

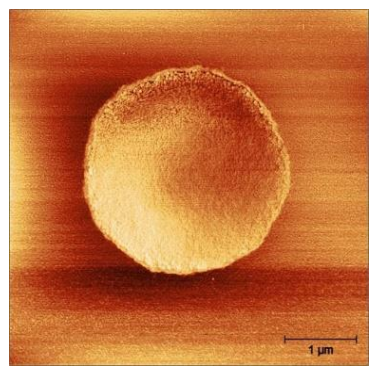

**Figure 7:** AFM phase images of selected (a,b) NIPAM-HMAM and (c,d) NIPMAM-HMAM microgels in the dry state, for different nominal compositions as indicated in the legend.

In Figure 8, selected phase images of the NIPMAM-HIPAM microgels are shown. More height and phase images for other compositions are shown in Figures S13 and S14. Focusing on individual particles in Figure 8, they show again a thin corona in all investigated cases, i.e. including the pure NIPMAM microgel. This corona is found to be independent of the comonomer content, while the apparent (flat) size increases slightly with comonomer content, as observed by DLS (see Figure S14e). The corresponding height profiles are shown in Figure S15. This is compatible with the relative size shown in Figure 4d, as shown in Figure S14(E), thus validating the comparison between 2D-AFM and 3D-DLS size analysis.

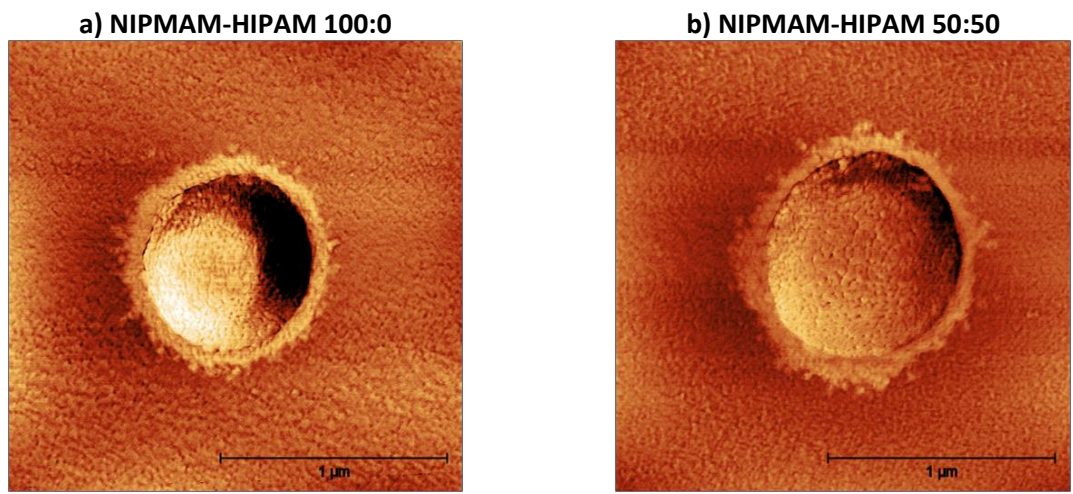

**Figure 8:** AFM phase images of selected NIPMAM-HIPAM microgels in the dry state, for nominal compositions increasing from (a) 0 mol% HIPAM, to (b) 50 mol%.

Figure 9, finally, shows phase images of NIPAM-HIPAM microgels with increasing HIPAM content. These microgels are the ones with the largest swelling, and their morphology is found to be different from the one of the other three systems. The dramatic increase in size with increasing hydrophilic comonomer content can be clearly seen in the sequence of Figures 9a-c. The microgels without HIPAM have a homogeneous circular shape with the already-mentioned thin corona around a denser core. When the comonomer content increases, the corona becomes larger and fuzzier. For the highest HIPAM incorporation, the corona extends over ca. 40% of the total microgel diameter. The corona thus covers some 80% of the particle volume. This might explain the high volumetric swelling ratio reported in Figure 6 for these systems, as it is plausible that this corona collapses completely to reach the rather small high-temperature hydrodynamic particle radius in suspension as reported in Figure 4b.

In Figure 9c, the phase lag allows to clearly identify on the 2D AFM picture the radial thickness of the corona of the 3D-particle, without however giving any information on the density of the corona. We have therefore attempted to further study the corona of the NIPAM-HIPAM (50:50) microgels by cross sectional height profiles. On a wafer examining flattened particles, the corona is found to give very little signal with respect to the microgel core, making it difficult to distinguish the corona from the background. Here, we propose an original approach: In Figure 9d, we have plotted first two height profiles across a particle center, which allows highlighting the overall particle size. The two cuts, along x and y, superimpose quite well, indicating that the particles are rather isotropic. Indeed, the difference in lateral extension is about 14%. In the inset, the corresponding cuts are illustrated by the horizontal and vertical lines across the microgel particle center.

In these height profiles along x and y, it appears to be impossible to identify the corona. Height sections along other lines, on the periphery of the core as indicated in the inset, have therefore been tested, and the results are also shown in Figure 9d. These profiles provide an indication of the height which can be considered "corona", to be identified by the maximum in the middle. We find 8 nm for the thickness of the corona in the dried and flattened case. By reporting this height on the above-mentioned first profiles crossing the entire particles, and identifying where the center-profile "hits" the corona thickness, we find an approximate corona thickness of 400 nm, for particles of core radius 635 nm. This is compatible with the above estimation from phase pictures showing that the size of the corona is about 40% of the total radius.

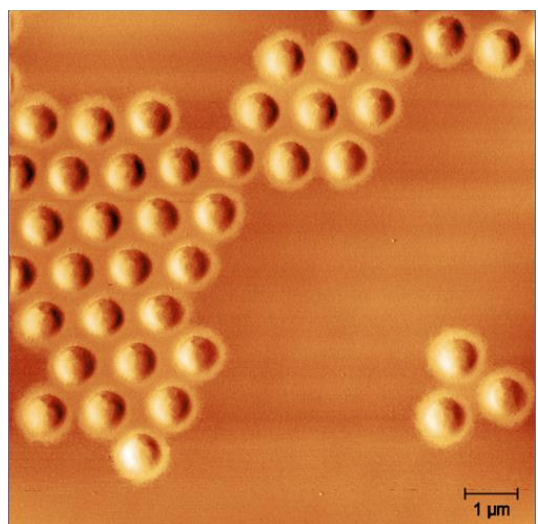

a) pure NIPAM

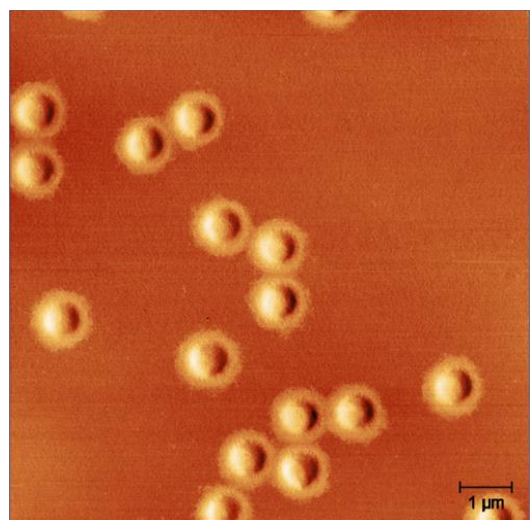

b) NIPAM-HIPAM 80:20

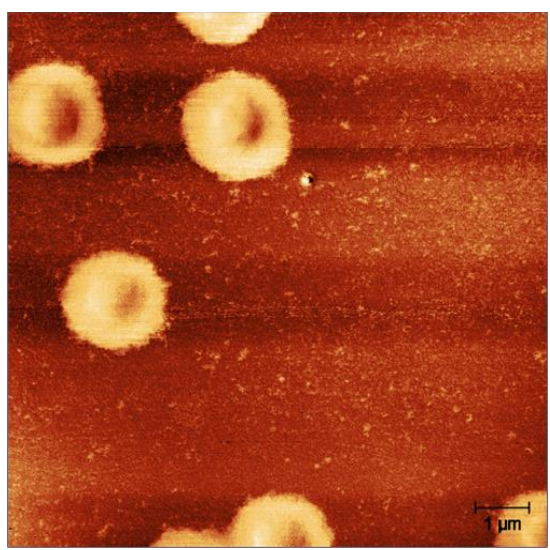

c) NIPAM-HIPAM 50:50

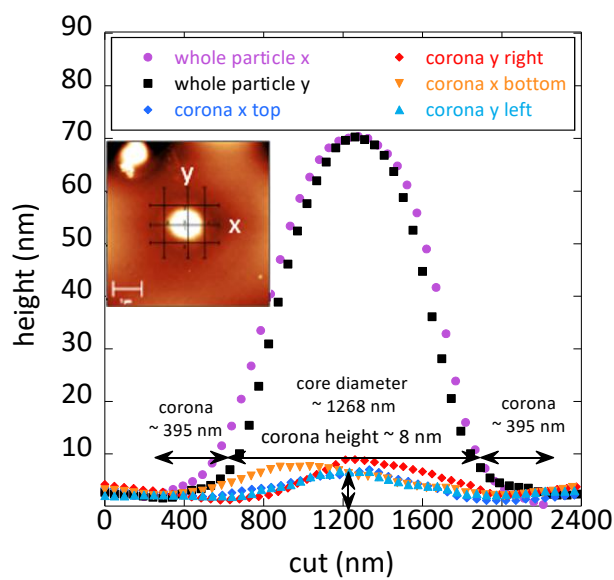

d) AFM height profiles of NIPAM-HIPAM 50:50

**Figure 9:** AFM phase images of (a) pure NIPAM, (b) 20mol% NIPAM-HIPAM, (c) 50 mol% NIPAM-HIPAM microgels in the dry state (nominal composition). d) Cross sectional heights of a 50:50 microgel according to the cuts indicated in the inset.

The remaining question pertains to the internal density of the corona. From Figure 9d, the outer, corona part is seen to result in a very low height, considerably lower than the height in the center of the core. When the particle is squeezed onto the surface by capillary forces, both the core and the corona are deformed, and it is difficult to separate their contributions to the height profile. In Figure S15, we have superimposed renormalized height profiles across the center of a "core-only" microgel (i.e., NIPAM-HIPAM 50:50 having only a very thin corona as shown in Figure 8b) to the one of a NIPAM-HIPAM (50:50) particle with corona. The NIPMAM particles have also been rescaled in their lateral extension to compensate for the size difference. The weak influence of the corona is found in the periphery of the particle. It is found to be asymmetric, because the underlying AFM snapshot caught a slightly asymmetric particle shape. Nonetheless, there is a clear "overintensity" in the peripheral region, which we attribute to the corona. This additional height of typically 0.1 of the total normalized height corresponds well to the 8 nm corona height identified in Figure 9d. These data suggest that the corona is about 10 times less dense than the core.

The results presented above show that the hydrophilic comonomers HIPAM and HMAM – which do not form microgels on their own in a conventional precipitation polymerization – could be successfully incorporated into NIPAM (resp. NIPMAM) microgels. The average incorporation has been determined by IR and NMR, and the resulting shift in VPTT as observed by cloud point measurements proves that copolymer microgels have been synthesized. The shift in VPTT, which has also been measured by DLS, could be extrapolated to the VPTT of the hypothetical pure HIPAM-BIS (resp. HMAM-BIS) particle, which we were unable to synthesize by precipitation polymerization in spite of several attempts. These VPTT values indicate a rather high transition temperature for hypothetical "pure" HMAM microgels of 99±1.5°C, and 67.5±0.5°C for HIPAM. These temperatures may serve as indicators of the "high VPTT properties" of these more hydrophilic molecules.

All microgel particles studied here show a moderate tendency to grow at low hydrophilic comonomer contents. For hydrophobic and non-thermosensitive comonomers such as isopropyl methacrylate, the opposite behavior has been observed in the literature.[61] As already discussed, Wedel et al. have found a growth in size correlated with higher VPTTs.[56] PEG-based microgels appear to follow the same trend: incorporating higher quantities of OEGMA into pMeO$_2$MA resulted in a VPTT shift by some 15 K, together with an increase in hydrodynamic radius from ca. 150 to 225 nm in the swollen state in water.[57] In this case, the formation of a fuzzy shell (or corona) at low temperatures is also reported. To summarize, these studies thus suggest that increasing hydrophilicity leads to bigger particles, as observed here.

Above a critical molar concentration, the particles studied here are found to increase considerably in size. This threshold is about 35 mol% for NIPAM-HIPAM and NIPMAM-HMAM, and 15 mol% for NIPAM-HMAM. For NIPMAM-HIPAM, the resulting comonomer contents were too low, but the data would also be compatible with a jump at 35 mol%. To the best of our knowledge, such a strong increase in size with composition has never been reported in the literature. We have also noted in the discussion of Figure 4 that at high comonomer incorporations, the swelling curves tend to be smoothed, not showing a very clear VPT anymore. This is qualitatively similar to what has been observed in the above-mentioned system with core-shell microgels, with the cores made of a monomer with VPTT (pNNPAM) lower than the one of the coronas (pNIPMAM).[68,69] In contrary to the present study, both monomers are microgel-forming on their own. A linear swelling was observed between the two transition temperatures. This linear swelling is akin to the smoothed transition observed in Figure 4 at high comonomer content, and it has later been explained by the gradual interpenetration of the two polymer networks.[69,86] It is very well possible that the resulting composition gradient in the present study is similar, thereby explaining the smoothing of the transition. On the other hand, a polydispersity in composition could lead to the same effect. A future SANS study with isotopic labelling might answer this question.

The issue of the physical mechanism of the behavior observed with increasing comonomer content, namely the first moderate growth in size, followed by a jump in total hydrodynamic radius, as well as the formation of a pronounced and increasing corona for the NIPAM-HIPAM system accompanied by strong (de)swelling can be discussed in the light of two possible origins. First, as already hypothesized by Wedel et al. [56], the decreasing fraction of more hydrophobic comonomer during synthesis – NIPAM or NIPMAM in our case – may lead to a lower number of nucleation sites at high temperature. In a simple model calculation, 30% less of the main monomer corresponding to a composition just below the size jump might correspond to 30% less nuclei, each having roughly 30% more available monomers, corresponding to an estimated increase of the radius of about 10 %. Such an argument would hold for homogeneous particles, and corresponds to the order of magnitude of the observed size increase for low incorporation of some 10 or 20 % in the collapsed state (Figure 4). It also corresponds roughly to the first, moderate increase in Figure 5, which focuses on the low temperature sizes, or to Figure S14E.

The strong increase in size, however, must be related to a change in molecular structure, which has been shown to become inhomogeneous by AFM (Figure 9). Given the hydrophilic nature of the additional comonomer, one may conjecture that these monomers participate only partially in the formation of nucleation sites. In this hypothesis, a natural composition gradient would arise, with the more hydrophobic main monomers dominating the center of the particles, surrounded by a more and more hydrophilic shell – neutron scattering might offer a possibility to check such composition gradients.[87,88] The presence of the growing corona in AFM in Figure 9 is fully compatible with this concept. The BIS-crosslinking would be most efficient in the center, leading to a higher modulus probed by AFM. The crosslinking density then decreases from the inner towards the peripheral region of the microgel, leading to a less and less densely crosslinked corona, as identified by AFM. It is unclear, however, why such a neat separation is observed in AFM. In any event, such a high-VPTT corona possesses a possibly very high swelling, as observed for NIPAM-HIPAM in Figure 6. This process would also explain why it is not possible to incorporate arbitrarily high comonomer contents, and in particular why pure HIPAM (resp. HMAM) microgels cannot be obtained: these molecules do not precipitate enough onto the nuclei at the synthesis temperature of 70 or 80°C. Just below the maximum incorporation, such microgels experience a strong growth, either of the more hydrophilic (and strongly deswellable) corona if it exists, or of the homogenous particle with high hydrophilic incorporation. The jump in size could thus be seen as the precursor for the impossibility to form microgels at higher comonomer contents.

## Conclusion

The swelling behavior and the microstructure of microgels copolymerized with two different high-VPTT monomers have been studied, evidencing a peculiar behavior as a function of comonomer content: at first, a moderate increase in size is observed, followed by a jump in size, and a usually less marked thermosensitivity with a broader VPTT. At the same time, the VPTT increases as expected, and it can be extrapolated to about 99°C for hypothetical pure HMAM microgels, and about 68°C for HIPAM. Although there is no guarantee that the linear law holds outside the measured range of concentrations, the knowledge of this number allows the simple calculation of any VPTT within the measured range. A third comonomer, NIPMAMol, is currently investigated for comparison. The microstructure has been investigated by AFM and mostly homogeneous spheres with possibly a small corona of dangling ends have been found. In one case, for NIPAM-HIPAM, which contrarily to all others shows a strongly increasing maximum swelling with increasing HIPAM content, AFM has detected a thick corona. The latter seems to grow with comonomer content and is probably responsible for the strong volumetric swelling of up to a factor of 100.

It would be highly interesting to investigate the detailed molecular structure, and particularly the monomeric density gradients of these microgels. Such data can be measured by SANS using isotopic substitution, and these experiments are currently planned. Note that the systems studied here have not been synthesized with deuterated monomers creating contrast for neutron scattering. In this context, the comparison to core-shell microgels which have been synthesized in a two-step procedure will be particularly insightful.

In conclusion, this study pushes precipitation-polymerization to its limits, due to the hydrophilic nature of the comonomers. Possibly synthesis at even higher temperature, and under pressure, would open pathways to new structures. It might also be worthwhile to explore alternative synthesis approaches like e.g. micro- or mini-emulsion polymerization with HMAM and HIPAM.[89,90] In the present article, we have outlined the possibilities of maximum incorporation, and described intriguing core-corona structures capable of extremely strong swelling. In all cases, the microgel VPTT could be shifted to higher temperatures, thereby showing the way – and the difficulties – of such high-VPTT microgel synthesis for energy applications, such as fuel cells.

**Supporting information:** HIPAM synthesis protocol, precipitation polymerization - synthesis parameters and methods, comonomer quantification – FTIR and NMR results, % transmittance of copolymers, DLS correlograms analysis, AFM analysis.

**Acknowledgements:** The authors are grateful for funding of the joint 'SmartBrane' project to the DFG (grant number HE 2995/14-1) and the ANR (grant number ANR-22-CE92-0052-01). Jean-Louis Bantignies and David Maurin are thanked for the FTIR measurements. Marco Wißbrock is thanked for NMR measurements.

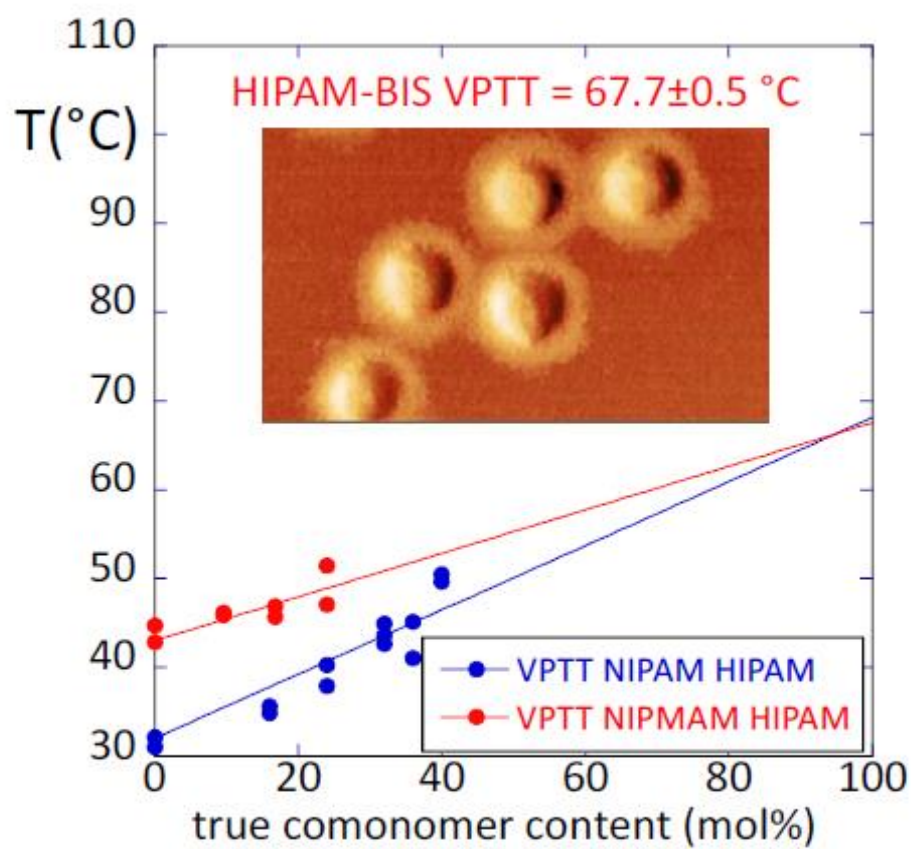

**Table of contents graphics.**

# Smart copolymer microgels with high volume phase transition temperature: Composition, swelling, and morphology

Aditi Gujare[1,2$], Stefanie Uredat[1,2$], Jonas Runge[1,2], Felix Morgenstern[2], Domenico Truzzolillo[1], Thomas Hellweg[2]*, Julian Oberdisse[1]*

[1] Laboratoire Charles Coulomb (L2C), University of Montpellier, CNRS, 34095 Montpellier, France.

[2] Department of Physical and Biophysical Chemistry, Bielefeld University, Universitätsstr. 25, 33615 Bielefeld, Germany

* Authors for correspondence: julian.oberdisse@umontpellier.fr, thomas.hellweg@uni-bielefeld.de

$ These Authors have contributed equally to this article.

HIPAM synthesis protocol, precipitation polymerization - synthesis parameters and methods, comonomer quantification – FTIR and NMR results, % transmittance of copolymers, DLS correlograms analysis, AFM analysis

## 1) HIPAM monomer synthesis

This monomer has been synthesized following the protocol published by Maeda et al.[1] The chemicals used were: 2-amino-1-propanol (Sigma-Aldrich, 98 %), acryloyl chloride (Thermosci., 96 %, 400ppm Phenothiazin), triethylamine (Acros, 99,7 %), chloroform (VWR, p.A.), 2-propanol (VWR, Normapur), cyclohexane (VWR, HIPERsolv), $D_2O$ (Roth, 99,8 %).

2-amino-1-propanol (0.1502 mol) and triethylamine (0.1520 mol) were dissolved in chloroform (150 ml) under $N_2$. This solution was cooled to -5 °C in an ice-NaCl-bath + acryloyl chloride (0.1495 mol) with stirring within 90 min. Stirring was continued for two hours at -5 °C. The chloroform was removed using a rotary evaporator + 2-propanol (180 ml). The solution was then stored at -20 °C overnight (~ 24 hours), colorless crystals were formed which were removed by suction filtration followed by rotary evaporation, leaving a yellow, viscous oil as the crude product. Further purification was carried out by column chromatography (EtAc EtAc/2-propanol mixture (4:1)). HIPAM was obtained as a colourless, highly viscous oil.

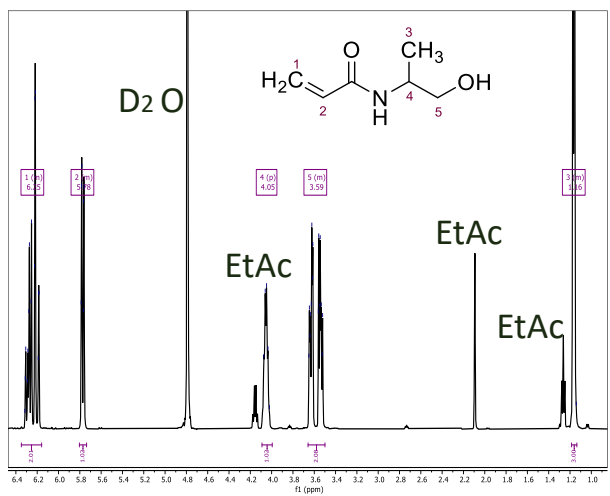

**Figure S1:** Characterization of HIPAM in $D_2O$ by $^1H$-NMR (Avance III 500, 500MHz): δ ($D_2O$, ppm) = 1.17 (d, 3H, C$\underline{H_3}$CHCH$_2$OH), 3.59 (m, 2H, CH$_3$CHC$\underline{H_2}$OH), 4.05 (m, 1H, CH$_3$C$\underline{H}$CH$_2$OH), 5.78 (d, 1H, CH$_2$=C$\underline{H}$), 6.25 (m, 2H, C$\underline{H_2}$=CH). Small impurities from EtAc δ ($D_2O$, ppm) = 1.26 (t, CH$_2$C$\underline{H_3}$), 2.09 (s, C$\underline{H_3}$CO), 4.14 (q, CH$_2$C$\underline{H_3}$)

## 2) Microgel synthesis – variations of protocol:

Two slightly different microgel synthesis routes have been used in the different labs. They are shown in Figures S2, S3 and S4 to produce equivalent results.

**Microgel synthesis method 1:** The reactions with HIPAM were carried out using recrystallized NIPAM (BLDpharm, 99.82 %, recrystallized in n-Hexane) and NIPMAM (TCI, > 98.0%, recrystallized in n-Hexane). The inert environment was maintained by continuous nitrogen purging, throughout the reaction. The reaction was stirred between 80 and 90°C for 4 hours followed by overnight stirring at room temperature. The suspension was filtered through glass wool and further purified by 5 successive centrifugations at 300,000xg for 30 minutes. This was then purified by 5 successive centrifugations at 300,000xg for 30 minutes. The crosslinker (BIS) concentration was 5 mol% of the total monomer content. The persulfate initiator used was ammonium persulfate (APS). Distilled water (Arium Pro VF, Sartorius Stedim) was used throughout the synthesis and the purification steps. The relevant parameters are summarized in Table S1.

**Microgel synthesis method 2:** The reactions were carried out using NIPAM (Sigma-Aldrich, >99 %) and NIPMAM (Sigma Aldrich, 97%), neither of them recrystallized in hexane. The inert environment was maintained by degassing distilled water with argon. The reaction was stirred at 90°C for 2 hours followed by overnight stirring at room temperature. This was then purified by three successive centrifugations at 18,000 x g for 60 minutes. The crosslinker (BIS) concentration was 5.4 mol% of the total monomer content. The persulfate initiator used was potassium persulfate. Distilled water (MilliQ water, Synergy® Water Purification System) was used throughout the synthesis and the purification steps. The relevant parameters are summarized in Table S1.

| System | NIPAM-HMAM | NIPMAM-HMAM | NIPAM-HIPAM | NIPMAM -HIPAM |
|---|---|---|---|---|
| Synthesis method | 2 | 2 | 1 | 1 |
| Temperature | 90°C 2h degassed | 90°C 2h degassed | 80-90°C 4h N$_2$ | 80-90°C 4h N$_2$ |
| Initiator | KPS | KPS | APS | APS |
| Max incorporation | 30mol% | 50 mol% | 50 mol% | 60 mol% |

|  | (40 mol% unsuccessful) | (60 mol% unsuccessful) | (70 mol% unsuccessful) | (70 mol% unsuccessful) |
|---|---|---|---|---|
| **BIS** | 5.4 mol% | 5.4 mol% | 5 mol% | 5 mol% |
| **Main monomers** | Non-recrystallized | Non-recrystallized | recrystallized | recrystallized |

**Table S1**: Synthesis parameters of four systems. All surfactant-free.

## 3) Nominal monomer contents in synthesis

The nominal monomer contents used in the microgel synthesis are summarized in Tables S2 and S3.

| System | Main monomer: NIPAM/NIPMAM | | | Comonomer: HIPAM | | |
|---|---|---|---|---|---|---|
| | Mol% | Millimoles | Mass (g) | Mol% | Millimoles | Mass (g) |
| NIPAM-HIPAM | 100 | 3.294 | 0.373 | 0 | 0 | 0 |
| | 80 | 2.72 | 0.308 | 20 | 0.683 | 0.088 |
| | 70 | 2.324 | 0.263 | 30 | 1.029 | 0.129 |
| | 60 | 2.023 | 0.229 | 40 | 1.365 | 0.176 |
| | 50 | 1.628 | 0.184 | 50 | 1.630 | 0.210 |
| NIPMAM-HIPAM | 100 | 2.987 | 0.380 | 0 | 0 | 0 |
| | 80 | 2.390 | 0.304 | 20 | 0.598 | 0.077 |
| | 65 | 1.942 | 0.247 | 35 | 1.046 | 0.135 |
| | 50 | 1.494 | 0.190 | 50 | 1.494 | 0.193 |

**Table S2:** HIPAM-based microgels monomer compositions (nominal) in ~1 wt. % reaction mixture (25 or 50 ml water).

| System | Main monomer: NIPAM/ NIPMAM | | | Comonomer: HMAM | | |
|---|---|---|---|---|---|---|
| | Mol% | Millimoles | Mass (g) | Mol% | Millimoles | Mass (g) |
| NIPAM-HMAM | 100 | 5.50 | 0.622 | 0 | 0 | 0 |
| | 90 | 4.95 | 0.560 | 10 | 0.55 | 0.056 |
| | 80 | 4.40 | 0.498 | 20 | 1.10 | 0.111 |
| | 70 | 3.85 | 0.436 | 30 | 1.65 | 0.167 |
| NIPMAM-HMAM | 100 | 5.50 | 0.699 | 0 | 0 | 0 |
| | 90 | 4.95 | 0.630 | 10 | 0.55 | 0.056 |
| | 80 | 4.40 | 0.560 | 20 | 1.10 | 0.111 |

| | 70 | 3.85 | 0.490 | 30 | 1.65 | 0.167 |
|---|----|------|-------|----|------|-------|
| | 60 | 3.30 | 0.420 | 40 | 2.20 | 0.223 |
| | 50 | 2.75 | 0.350 | 50 | 2.75 | 0.278 |

**Table S3**: HMAM-based microgels monomer compositions (nominal) in ~1 wt. % reaction mixture (75 ml water).

## 4) Influence of different protocols: method 1 and method 2

In Figure S2, we show the original and the normalized swelling curve of NIPAM-co-HMAM (monomer ratio 90:10) microgels synthesized by the two different methods. In addition, we have exchanged the initiator of method one to KPS in order to check the influence of the two different protocols. Comparable size and VPTTs are obtained by the two different methods.

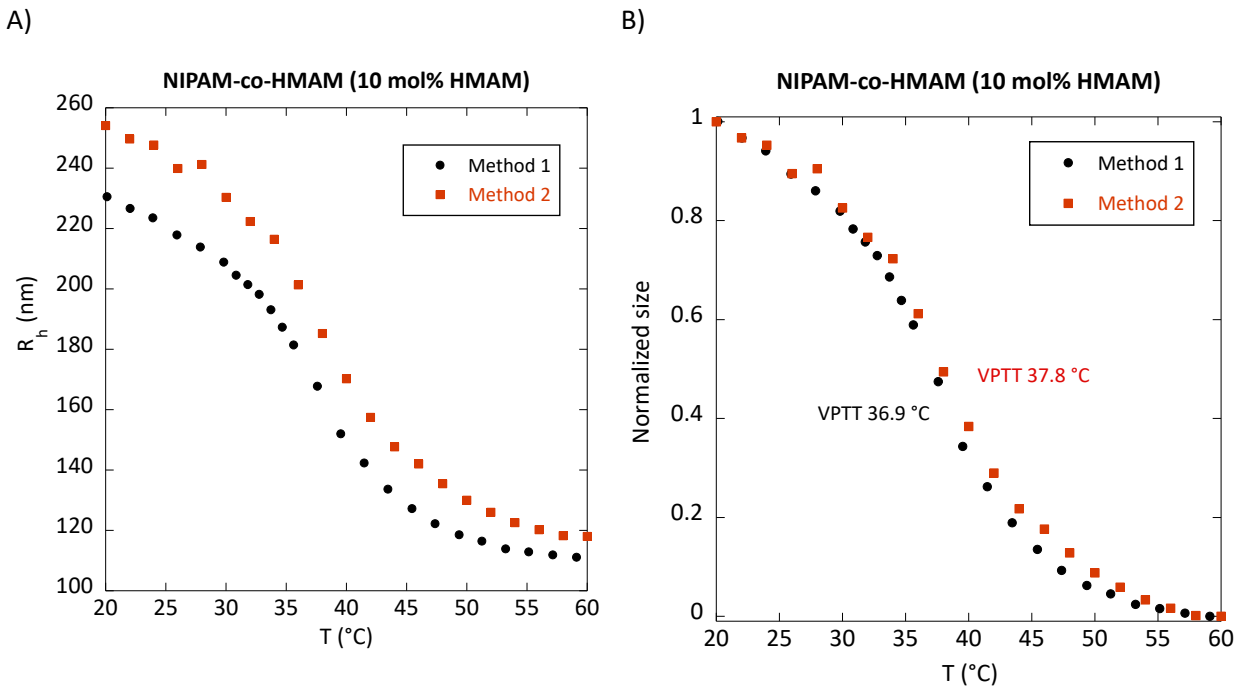

**Figure S2:** (a) Swelling curves of a NIPAM-HMAM (90:10) microgel synthesized following the two different protocols. (b) Same curves normalized in hydrodynamic radius between 1 and 0 in order to highlight the very low VPTT change from 36.9 to 37.8 °C

## 5) Influence of different initiators

Furthermore, the impact of the different initiators, KPS or APS, respectively, has been tested with pure (non-recrystallized) NIPMAM microgels synthesized by method 2. Although comparable VPTTs are observe in these microgels, different initiators might affect the size and the polydispersity of the microgels slightly.

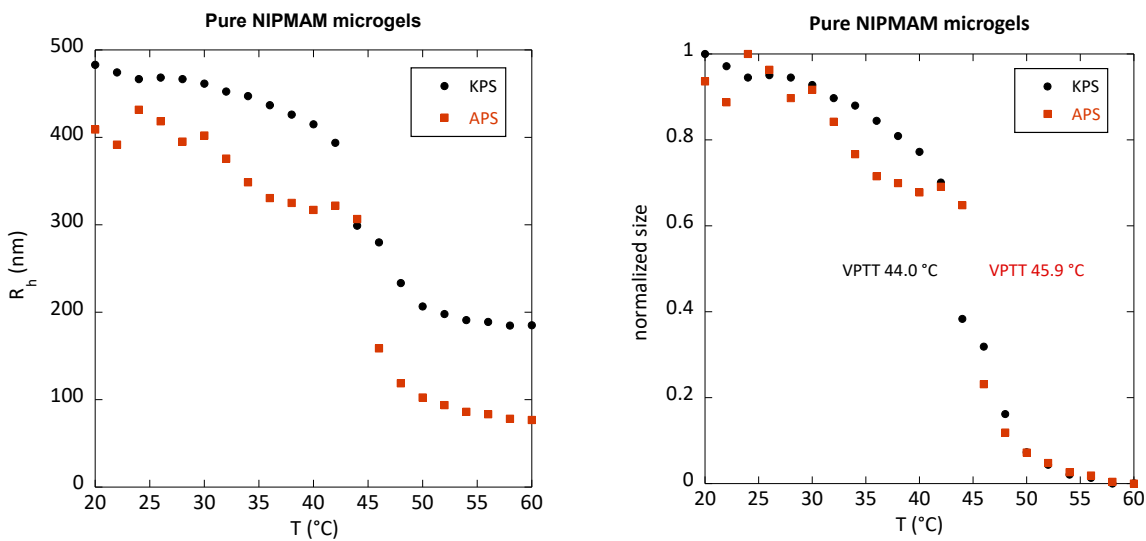

**Figure S3:** Swelling curves of a pure NIPMAM microgels synthesized by method 2, comparing the influence of different initiators, with VPTTs varying from 44.0 to 45.9°C.

In summary, **HIPAM-based microgels** were synthesized by method 1 using the initiator APS (ammonium persulfate) whereas the **HMAM-based microgels** were synthesized by method 2 using the initiator KPS (potassium persulfate), without any major impact on the resulting swelling curves of the microgels. The different synthesis methods are identified in Table S1, together with the maximum incorporation. With NIPAM, clots occurred at 70 mol% HIPAM, and a hard polymer layer was formed at 100 mol%, see Figure S5 as example for pure HIPAM. It is thus impossible to synthesise pure HIPAM-BIS microgels.

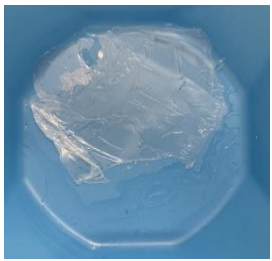

**Figure S4**: Microgel synthesis with 100 mol% HIPAM which resulted in a polymer layer and not in microgels.

## 6) Experimental determination of incorporation of comonomers into microgels

Figure S6 describes the incorporation rate of the comonomer HMAM showing the FTIR spectrum of the microgels with the alcohol stretching frequency (1034 cm-1) compared to the corresponding calibrations (known monomer compositions).

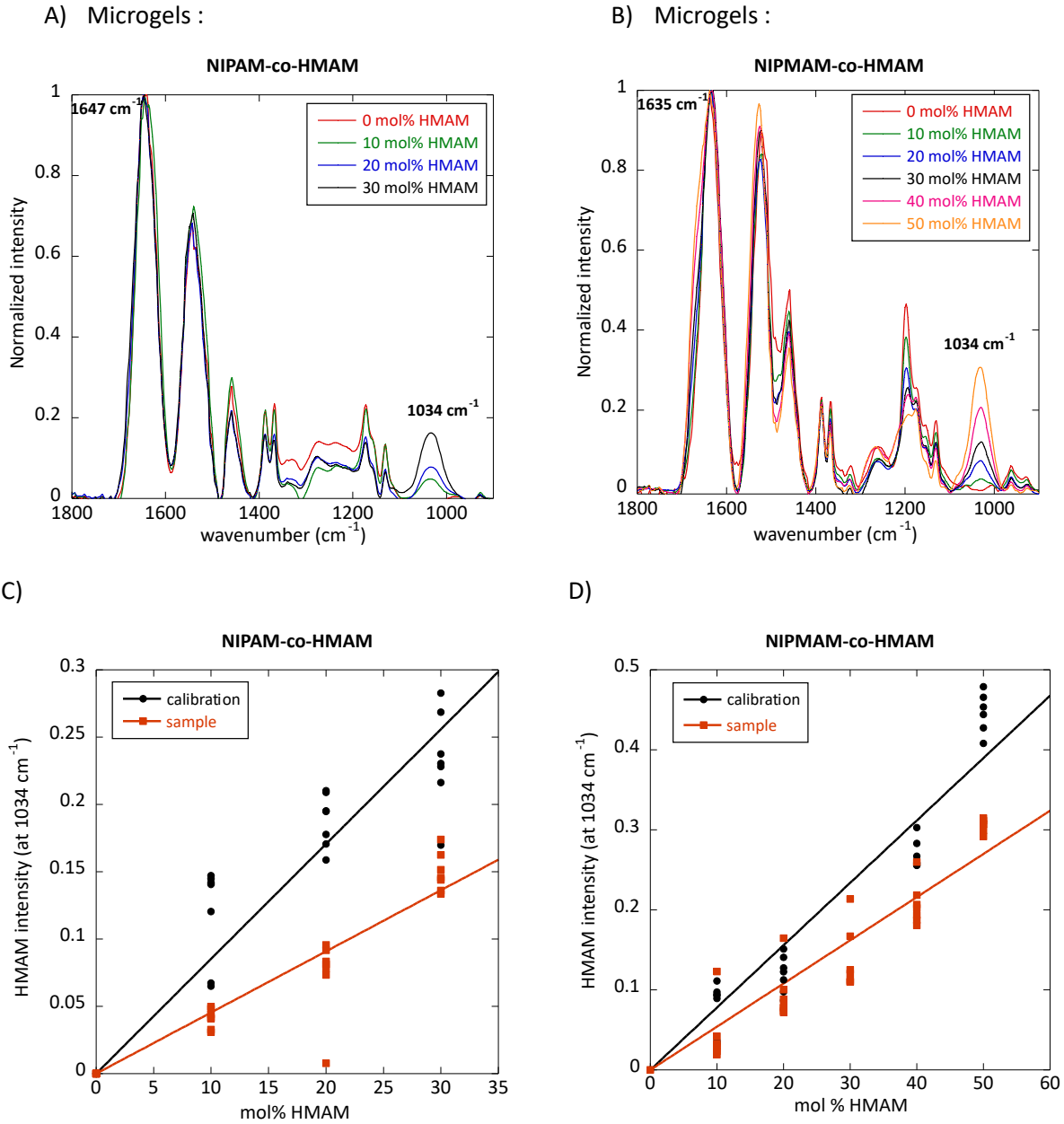

A) Microgels :

B) Microgels :

C)

D)

E)

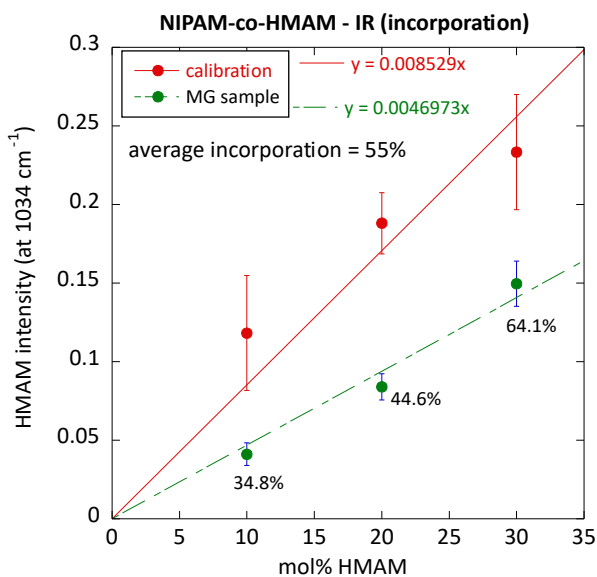

F)

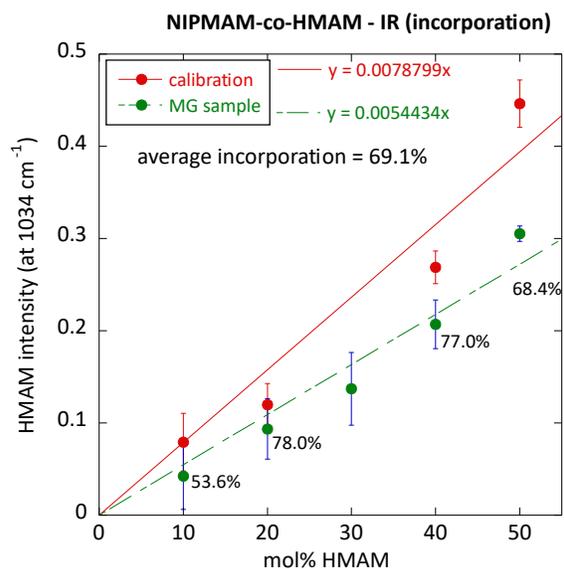

**Figure S5**: Normalized IR spectra of (A) NIPAM-co-HMAM, (B) NIPMAM-co-HMAM with their calibration curves (C) and (D), and their average incorporation (E) and (F) to determine the amount of the monomers and their ratios in the microgels.

HMAM is known to undergo condensation reaction upon thermal treatment. [2] It is therefore possible that the freeze-dried samples for performing IR in the ATR mode, could have some HMAM molecules that have undergone condensation, thus resulting in an underestimation of the intensity of the alcohol stretching frequency (1034 cm-1).

The IR measurements with HIPAM-microgels resulted in a double peak which impeded further analysis. Therefore proton NMR was performed. NMR peak height has been normed for visualization, but all calculations with NMR involve peak areas. The ratio between a signal where both monomers (NIPAM and HIPAM) are contributing to and a signal which is only appearing due to HIPAM (Figure S7).

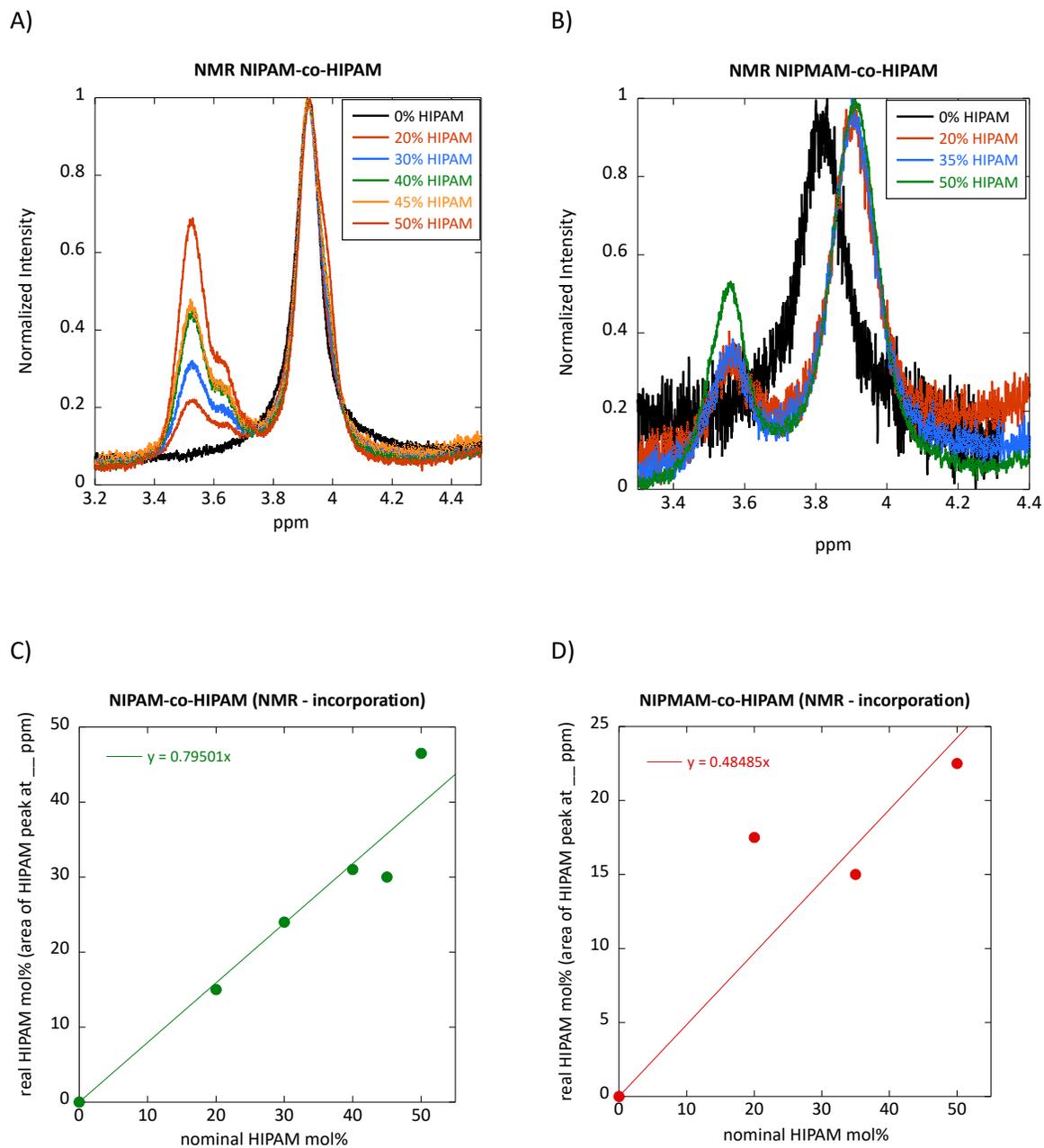

**Figure S6**: Normalized NMR spectra of (A) NIPAM-co-HIPAM, (B) NIPMAM-co-HIPAM resulting in HIPAM incorporation as in (C) and (D).

## 7) Cloud point measurements by turbidimetry

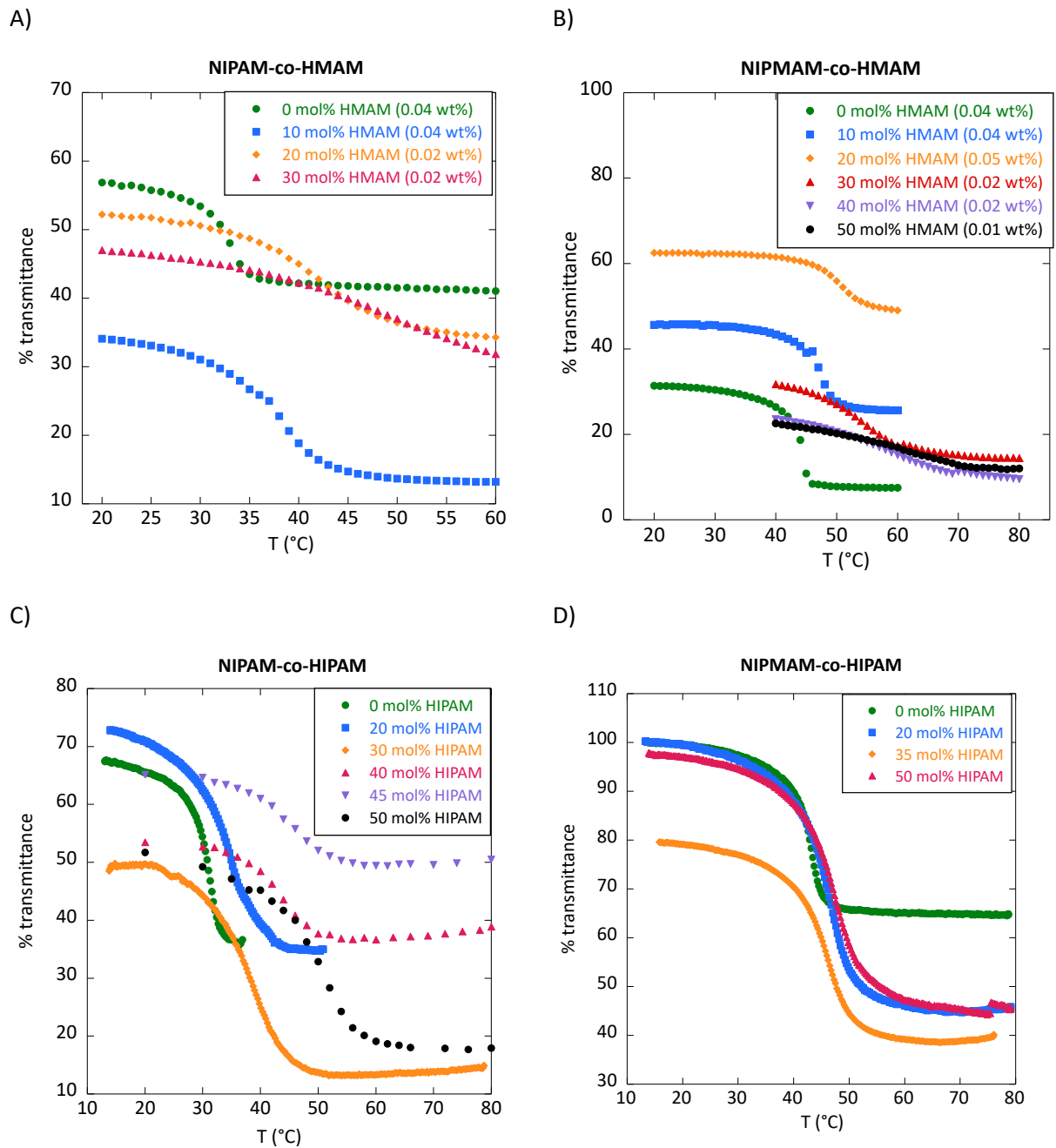

**Figure S7**: Turbidity measurements for the microgel systems: (A) NIPAM-co-HMAM (B) NIPMAM-co-HMAM, (C) NIPAM-co-HMAM and D) NIPMAM-co-HMAM microgels.

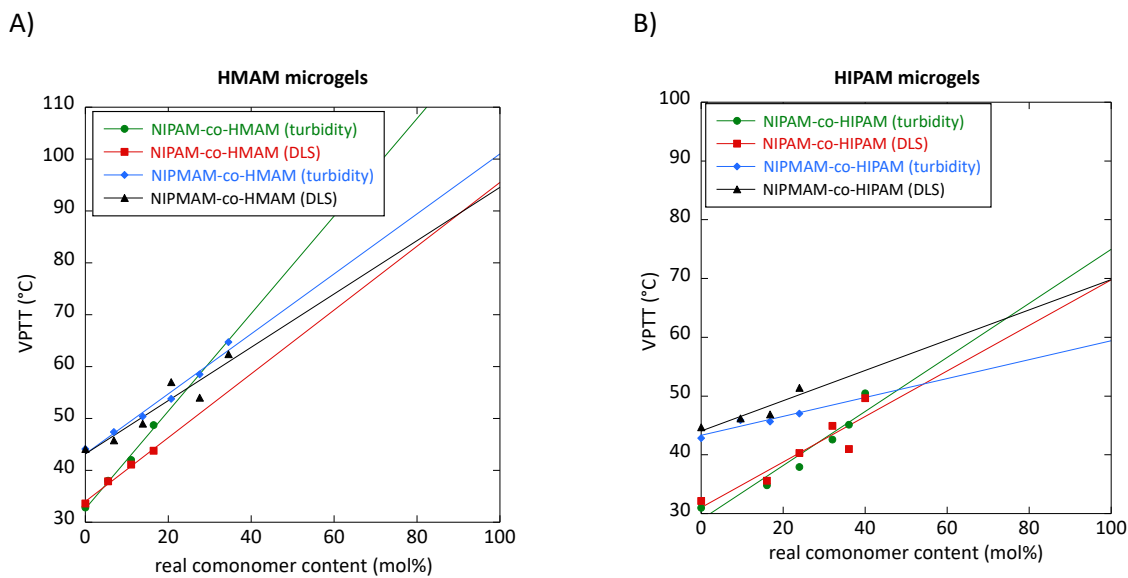

**Figure S8:** VPTT determined from turbidity and DLS for (A) HMAM microgels and (B) HIPAM microgels.

# 8) Analysis of noisy DLS correlograms

Below are two examples of noisy DLS data, where the average decay is the result of multiple decays of the auto-correlation function arising due to high polydispersity in hydrophilic systems. The corresponding hydrodynamic radius as a result of the average decay (in red) versus the hydrodynamic radius as a result of the first decay (in black) in polydispersed systems:

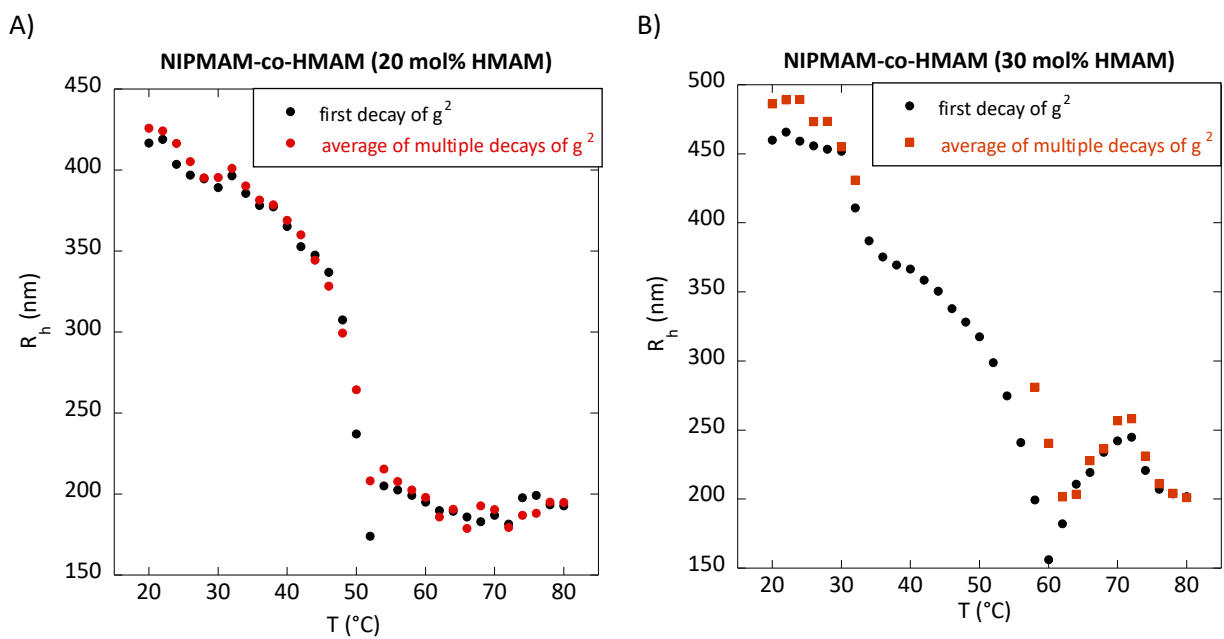

**Figure S9:** $R_h$ vs T from DLS for NIPMAM-co-HMAM systems with A) 20 and B) 30 mol% HMAM.

An example of a monodispersed versus a polydispersed correlation function, along with their intensity traces (NIPMAM-HMAM systems in the following examples):

A) Monodispersed system (10 mol% HMAM)    B) Polydispersed system (50 mol% HMAM)

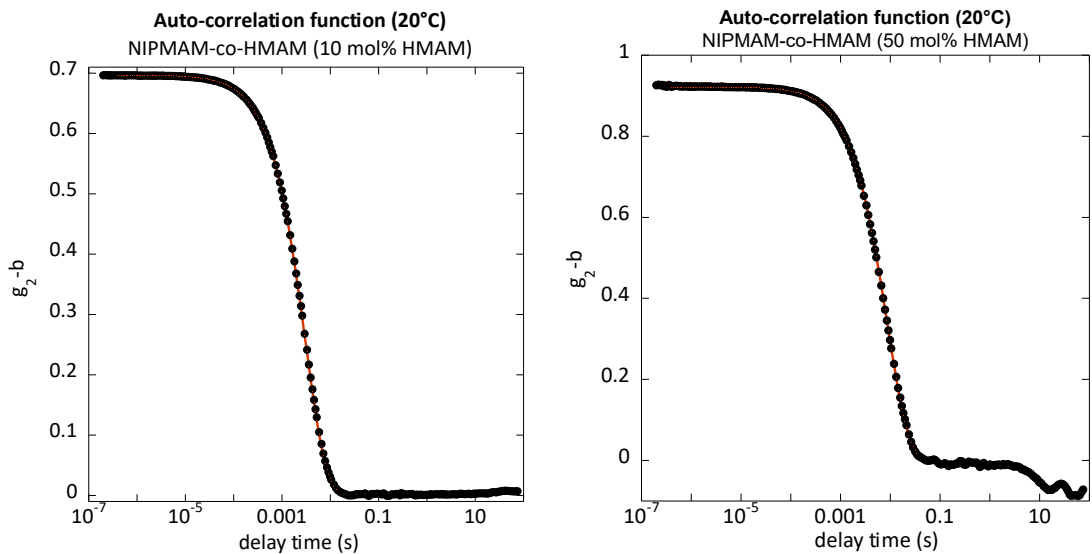

C) Monodispersed system (10 mol% HMAM)    D) Polydispersed system (50 mol% HMAM)

**Figure S10**: (A) Correlation function with a single decay confirming a monodispersed microgel system with (C) its corresponding intensity trace. (B) Correlation function that is noisy at larger lag times and (D) its corresponding intensity trace showing spikes describe a polydispersed microgel system. Such correlation functions are seen in microgels with higher hydrophilic comonomer content.

## 9) AFM results

Figure S11 and S12 are AFM images of the HMAM systems in the dry state, confirming the homogeneity (no core-shell structure) in all the HMAM-based microgels.

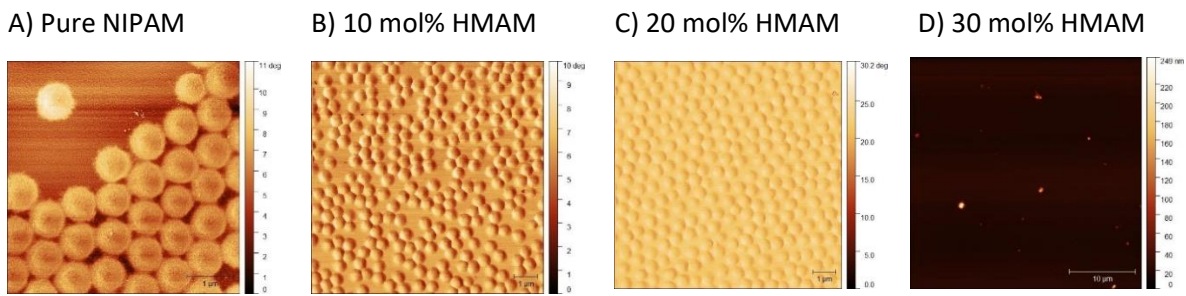

**Figure S11:** AFM images of NIPAM-co-HMAM homogeneous microgels (A) pure NIPAM, (B) 10 mol% HMAM, (C) 20 mol% HMAM, (D) 30 mol% HMAM

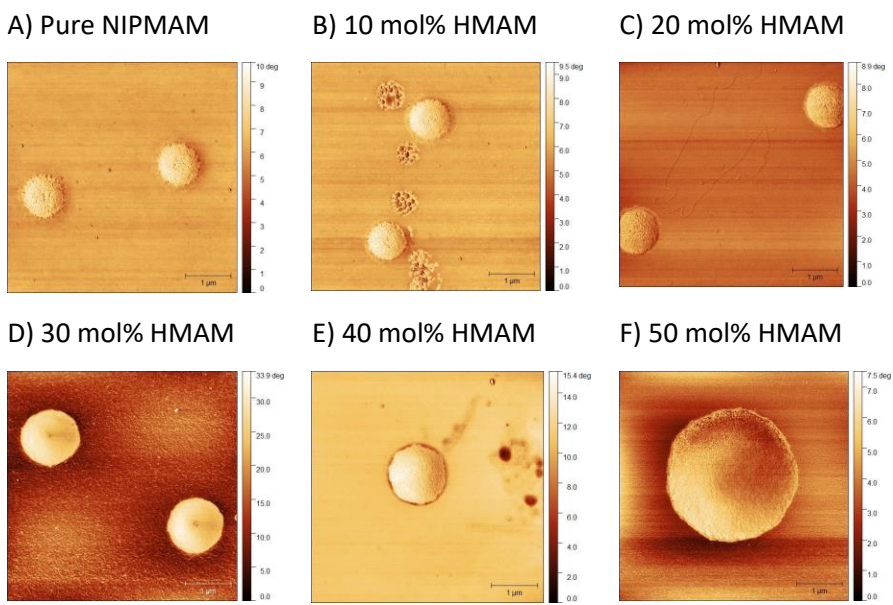

**Figure S12:** AFM images of NIPMAM-co-HMAM homogeneous microgels (A) pure NIPMAM, and nominal compositions of HMAM: (B) 10 mol% HMAM, (C) 20 mol% HMAM, (D) 30 mol% HMAM, (E) 40 mol% HMAM, (F) 50 mol% HMAM

Figure S13 and S14 are AFM images of the NIPMAM-HIPAM system which also have a homogeneous structure (show a thin corona). Figure S14 (E) shows the size change with increasing HIPAM content, as seen by AFM and DLS analysis.

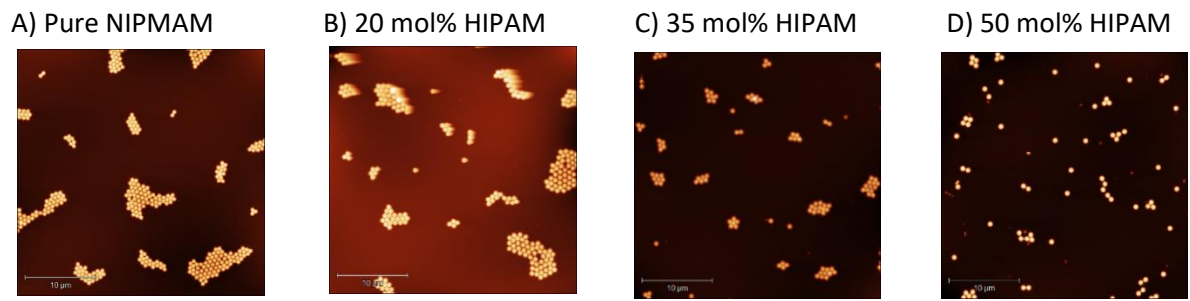

A) Pure NIPMAM    B) 20 mol% HIPAM    C) 35 mol% HIPAM    D) 50 mol% HIPAM

**Figure S13:** AFM height images of NIPMAM-HIPAM microgels in the dry state, for nominal compositions increasing from (a) 0 mol% HIPAM, to (b) 20 mol%, (c) 35 mol%, and (d) 50 mol%.

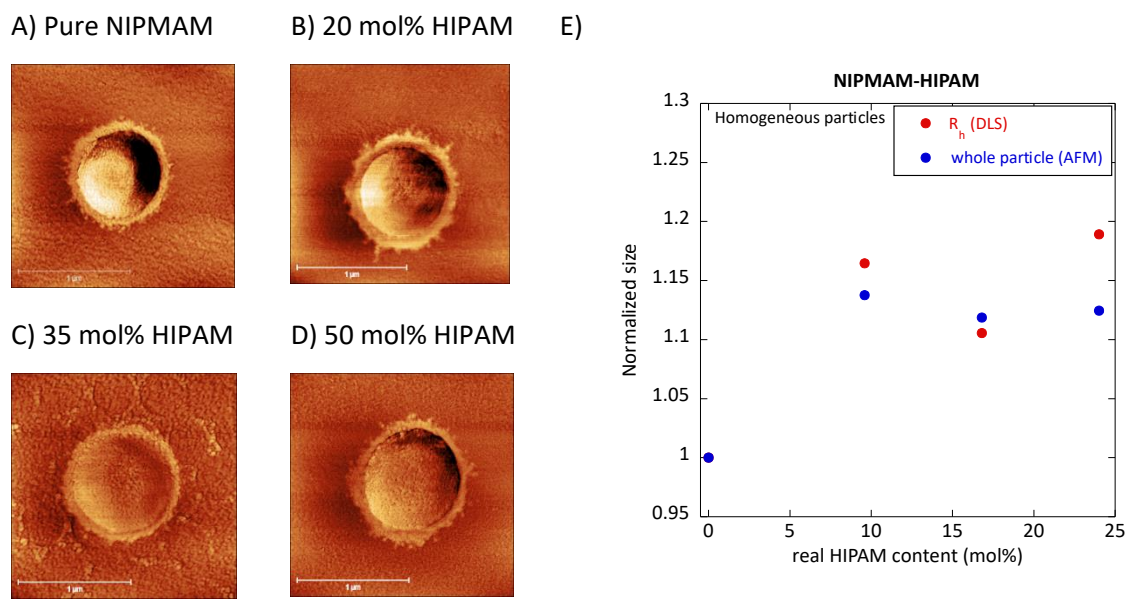

A) Pure NIPMAM    B) 20 mol% HIPAM    E)

C) 35 mol% HIPAM    D) 50 mol% HIPAM

**Figure S14:** (A-D) AFM phase images of NIPMAM-HIPAM microgels in the dry state, for nominal HIPAM compositions. (E) Comparison of the normalized size of the particle as seen in AFM to the hydrodynamic radius for increasing HIPAM composition in NIPMAM-HIPAM microgels at low temperature.

Comparison between homogeneous microgels and the coreshell microgels (Figure S15) as seen in the height profile:

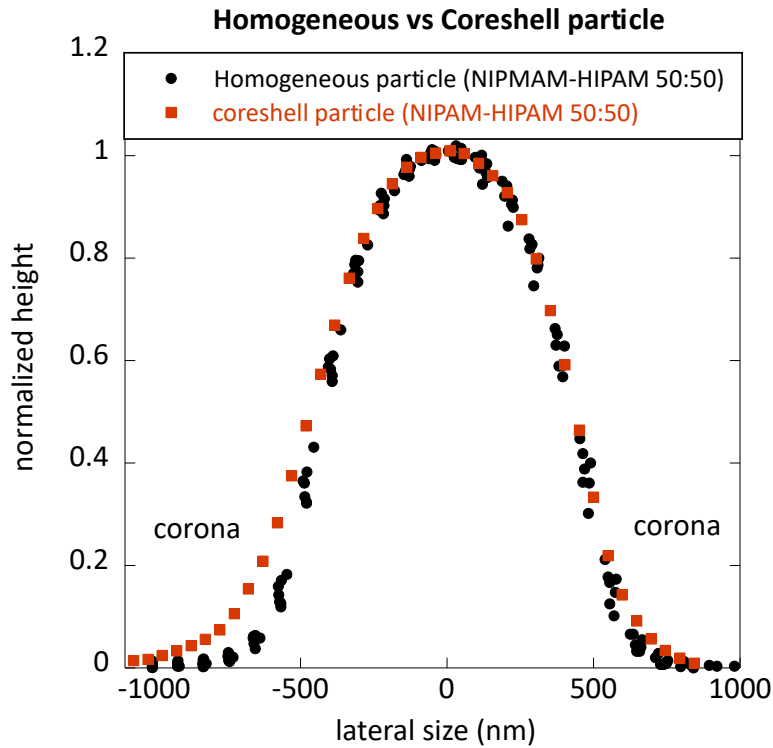

**Figure S15**: height profiles for NIPAM -HIPAM 50:50 core-corona microgel compared to a corona-free NIPMAM-HIPAM (50:50) microgel. For the latter seven cuts across different particles have been combined without rescaling, showing the robustness of the particle monodispersity. Both NIPMAM and NIPAM microgels have been rescaled to identical height and shape around the maximum.